\documentclass[preprint,onecolumn,amsmath,letterpaper,amssymb,prl,floatfix]{revtex4-1}

\usepackage{graphicx}
\usepackage[font=small, justification=centerlast]{caption}
\usepackage{color}
\usepackage{comment}
\usepackage{siunitx}
\usepackage{amsmath}

\usepackage[sort&compress]{natbib}

\begin{document}
%\bibliographystyle{nature1}

%\begin{comment}
\title{Discrete step sizes of molecular motors lead to bimodal non-Gaussian velocity distributions under force}

\author{Huong T. Vu$^1$, Shaon Chakrabarti$^1$, Michael Hinczewski$^2$, and D. Thirumalai$^1$} \affiliation{$^1$Biophysics Program, Institute For
  Physical Science and Technology, University of Maryland, College
  Park, MD 20742\\$^2$Department of Physics, Case Western Reserve University, Cleveland, OH, 44106}

\begin{abstract}
  Fluctuations in the physical properties of biological machines are
  inextricably linked to their functions.  Distributions of
  run-lengths and velocities of processive molecular motors, like
  kinesin-1, are  accessible through single molecule
  techniques, yet there is lack a rigorous theoretical model for these
  probabilities up to now.   We derive exact analytic results for a kinetic model to
  predict the resistive force ($F$) dependent velocity ($P(v)$) and
  run-length ($P(n)$) distribution functions of generic finitely
  processive molecular motors that take forward and backward steps on
  a track.  Our theory quantitatively explains the zero force
  kinesin-1 data for both $P(n)$ and $P(v)$ using the detachment rate
  as the only parameter, thus allowing us to obtain the variations of
  these quantities under load. At non-zero $F$, $P(v)$ is
  non-Gaussian, and is bimodal with peaks at positive and negative
  values of $v$.  The  prediction that $P(v)$ is  bimodal 
is a consequence of the discrete step-size of
  kinesin-1, and remains even when the step-size distribution is taken
  into account. Although the predictions are based on analyses of
  kinesin-1 data, our results are general and should hold for any
  processive motor, which walks on a track by taking discrete steps.
\end{abstract}

\date{\today}

\maketitle
\def\s{\rule{0in}{0.28in}}

Molecular motors convert chemical energy, typically from ATP
hydrolysis, into mechanical work to facilitate myriad activities in
the cell, which include gene replication, transcription, translation,
and cell division \cite{bookLodish,Svoboda1994,schliwa2003}. Despite
the bewildering variations in their sequences, structures, and
functions, a large number of cellular motors take multiple steps
directionally along linear tracks and are known as processive
motors. Well-known in this category are kinesins \cite{Ray1993},
myosins \cite{Spudich2010} and dyneins \cite{Roberts2013} that carry
vesicles and organelles along microtubules or actin filaments
\cite{bookHoward,Mallik2004}, and helicases that unwind nucleic acid
strands while translocating on them
\cite{Lohman2008,Singleton2007}. Fundamental insights into their
functions have emerged from single molecule experiments
\cite{Kojima1997,Nishiyama2002,Fischermeier2012,Cappello2006,Uemura2002,
  Svoboda1993,Svoboda1994,Fehr2008,Schnitzer2000,Milic14PNAS} combined
with discrete stochastic (or chemical kinetics) models
\cite{Kolomeisky2007,Schnitzer1995,Taniguchi2005,Shaevitz2005,Chowdhury2013}.
These studies have focused on quantities like mean velocity, mean run-length,
and the dwell-time distribution \cite{Fisher2001,Schnitzer2000}, at
varying external forces and ATP concentrations. However, much less
attention has been paid to distributions of motor velocities in
experiments or in theoretical models.  Given the inherently stochastic
nature of the motor cycle, velocity fluctuations must play an important
role in motor dynamics.  However, up to now  analytical
tools to interpret the fluctuation data,  readily available from
experiments, do not exist.  This letter seeks to address that gap, providing
universal closed-form expressions for velocity and run-length
distributions valid for any processive motor.

\begin{figure}
\centerline{\includegraphics[width=\columnwidth]{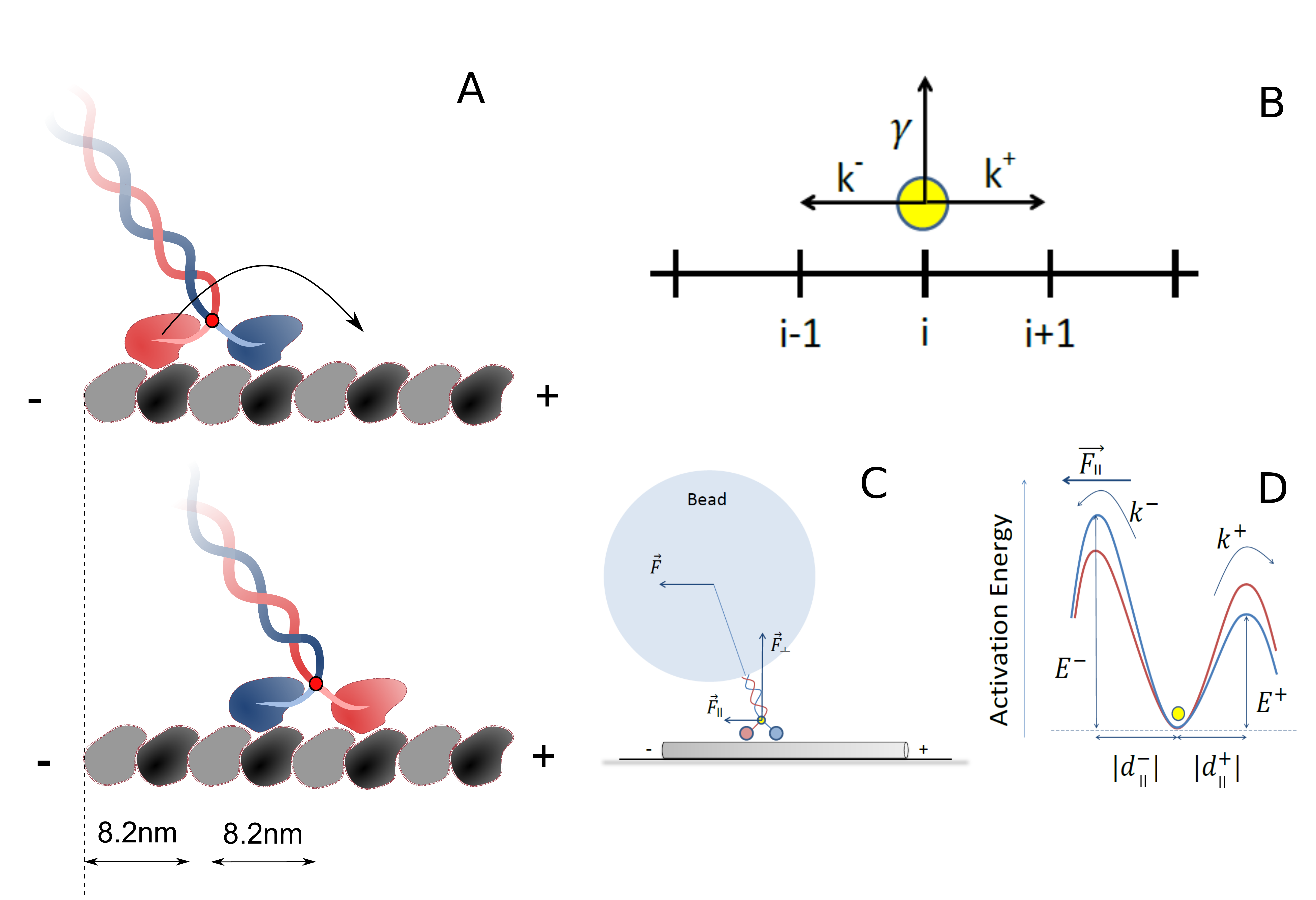}}
\caption{{\bf A:} Schematic of a kinesin molecule walking
  hand-over-hand on a microtubule (MT) with a discrete step-size of
  8.2~nm. {\bf B:} Sketch of the model in which the yellow circle
  represents the center-of-mass (COM) of the kinesin (capturing the
  point in red in {\bf A}). The position of the COM on the MT is
  denoted by $i$. Kinesin can step ahead, back, and detach from the
  microtubule with rates $k^+$, $k^-$, and $\gamma$ respectively. {\bf
    C:} Decomposition of the resistive optical trap force $F$ applied
  to the bead attached to the coiled-coil, along components parallel
  ($F_\parallel$) and perpendicular ($F_\perp$) to the MT axis
  \cite{Svoboda1994}.  {\bf D:} Energy landscape for forward and
  backward rates with the blue and the red curves corresponding to
  zero and non-zero $F$ respectively.  $F_\parallel$ increases the
  backward rate and decreases the forward rate; $F_\perp$ increases
  the detachment rate.}
\label{fig1}
\end{figure}

Let us assume a motor  takes steps of size $s$ on a polar track, and
moves a net displacement $n s$ before detaching at time $t$, where $n$
is an integer known as the run-length.  Then the natural definition
for average velocity for the trajectory is $v = ns/t$.  Our work
tackles several key questions about the distributions $P(v)$ and
$P(n)$ for molecular motors under force.  When can $P(v)$ be
represented by a Gaussian, an approximation used to analyze
experiments
\cite{Fischermeier2012,Ali2008,Soppina2014,Xu2013,Hammond2009,Cappello2006}?
If the detachment rate from the polar track
$\gamma$ is negligible compared to the forward rate $k^+$
(Fig~\ref{fig1}B), the motor walks a large number of steps forward
before detachment then $P(v)$ should approximately be a Gaussian, as
expected from the Central Limit Theorem (CLT). However, what is the
behavior of $P(v)$ and $P(n)$ when $\gamma$ becomes comparable or even
larger than the other rates involved, situations encountered in
single-molecule experiments in the presence of external force
(Fig~\ref{fig1}C \cite{Svoboda1994,bookdetachmentrate})? To address
these questions, we derive exact analytical expressions for $P(n)$ and
$P(v)$, using a simple but accurate kinetic model with only three rate
parameters (Fig~\ref{fig1}B).  The model has a broad scope, allowing
analysis and prediction of experimental outcomes for a large class of
processive motors and other rotary machines.

The central results of this work are: (i) At non-zero $F$, $P(v)$ is
non-Gaussian because $\gamma$ cannot be neglected compared to $k^+$,
resulting in the number of steps being not large enough for CLT to be
valid. Even when $F=0$ there is a discernible deviation from a
Gaussian distribution.  (ii) Surprisingly, when $F \ne 0$, $P(v)$ is
 asymmetric about $v=0$ with a bimodal shape containing peaks,
one at $v>0$ and the other at $v<0$. With increasing $F$, the peaks
become symmetrically positioned with respect to $v=0$ and completely
symmetric at the stall force $F_S$. (iii) As $F$ exceeds $F_S$,
reaching the superstall regime, the peak position at $v>0$ ($v<0$)
moves to higher (lower) values. These counter-intuitive results are
consequences of the discrete nature of steps that molecular motors
take on their tracks.

In our simplified model (Fig. \ref{fig1}B) (a model with intermediates is analyzed in the Supplementary Information), the motor can move towards the
plus/minus ends of the track or detach from each site. The
distribution of times that the motor stays attached to the track is
$P(t)=\gamma e^{-\gamma t}$.  The probability, $P(m,l)$, that the
motor takes $m$ forward, $l$ backward steps before detachment is,
\begin{equation}
P(m,l)=\left(\frac{k^+}{k_T}\right)^m\left(\frac{k^-}{k_T}\right)^l\left(\frac{\gamma}{k_T}\right)\frac{(m+l)!}{m! l!},
\label{2Pml}
\end{equation}
where $k_T=k^{+} +k^{-}+ \gamma$ is the total rate, $\frac{k^{+}}{k_T}$ ($\frac{k^{-}}{k_T}$)  is the probability of taking a forward (backward) step, and $\frac{\gamma}{k_T}$ is the detachment probability.  The final term in Eq.\ref{2Pml} accounts for the number of ways of realizing $m$ forward and $l$ backward steps.  The run-length distribution, $P(n)$, where $n=m-l$, is 
\begin{eqnarray}
P(n)&=&\sum_{m,l=0}^{\infty}\left(\frac{k^+}{k_T}\right)^m\left(\frac{k^-}{k_T}\right)^l\left(\frac{\gamma}{k_T}\right)\frac{(m+l)!}{m! l!}\delta_{(m-l),n}.
\end{eqnarray}
We find that  $P(n=0) = \frac{\gamma}
{\sqrt{k_T^2-4k^+k^-}}$ and $P(n\gtrless
 0)$  takes the simple form (see SI for details), 
\begin{eqnarray}
P(n\gtrless 0)&=&\left(\frac{2k^\pm}{k_T+\sqrt{k_T^2-4k^+k^-}}\right)^{\pm n}\frac{\gamma}
{\sqrt{k_T^2-4k^+k^-}}.
\label{2Pn}
\end{eqnarray}

The velocity distribution is given by
$P(v)=\sum_{n=-\infty}^\infty \int_{0}^\infty
dt\,\delta(v-n/t)P(n,t)$,
with distances measured in units of step-size $s$.  Here
$P(n,t)=\sum_{m,l=0}^\infty \delta_{(m-l),n}P(m,l,t)$, where the joint
distribution $P(m,l,t)$ for $m$ forward, and $l$ backward steps with
detachment at $t$ is (see SI for details),
\begin{equation}
P(m,l,t)=\frac{t^{m+l}}{m!l!}(k^+)^m(k^-)^l\gamma\exp(-k_Tt).
\label{2Pmlt}
\end{equation}

The exact expressions for $P(v\gtrless 0)$ are
\begin{eqnarray}
P(v\gtrless 0)=\frac{\gamma}{|v|}\sum_{n=0}^{\infty}\left(\frac{n}{|v|}\right)^{n+1}\frac{1}{ n!}\left(k^\pm e^{-\frac{k_T}{|v|}}\right)^n{}_0F_1\left(;n+1;\frac{n^2k^+k^-}{|v|^2}\right).
\label{2Pv>}
\end{eqnarray}

Our approach also allows us to compare $P(v)$ with the distribution of
$P(v_\text{inst})$ of ``instantaneous velocity'', an alternative
measure of motor dynamics.  If the dwell-time of a motor at a site is
$\tau$, then $v_\text{inst} = s/\tau$.  The distribution
$P(v_\text{inst})$, can in principle be computed from $P(\tau)$, the
distribution of dwell-times \cite{Valleriani08EPL,Tsygankov07PRE,Linden07BJ}. Analytical results for
$P(v_\text{inst})$ are discussed in the SI.

\begin{figure}
\centerline{\includegraphics[width=1\columnwidth]{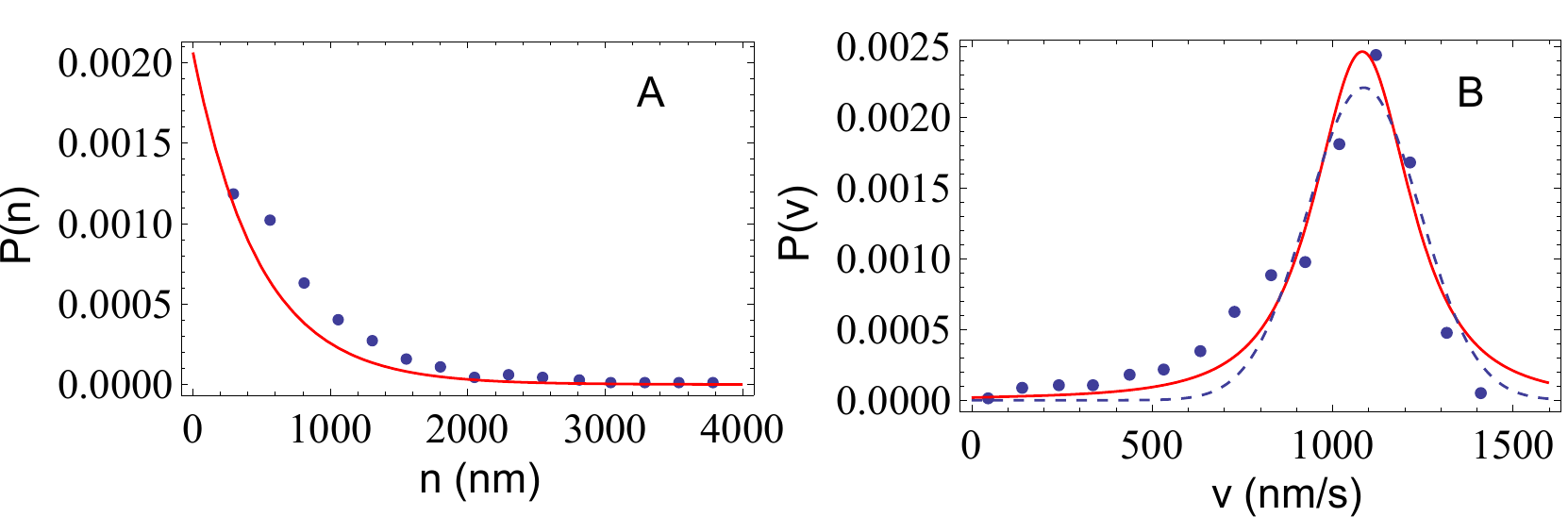}}
\caption{Simultaneous fits (red lines) of zero force Kin1 data (blue
  dots) \cite{Fischermeier2012} for run-length ({\bf A}:
  $P(n)$--Eq.~\ref{2Pn}) and velocity ({\bf B}:
  $P(v)$--Eq.~\ref{2Pv>}) distributions. The dashed line in {\bf B} is
  a Gaussian fit. It should be stressed that the results in {\bf (A)}
  and {\bf (B)} were fit using a single parameter, $\gamma_0$, with
  the extracted zero force values for $k^{+}_0$ and $k^{-}_0$ in
  Table~\ref{tablerates}(a).}
\label{figwalterdata}
\end{figure}

\begin{table}[t]
\centering
\caption{Force-dependent rates for Kin1. (a) and (b) are rates at $F=0$ obtained from experimental data on Kin1/acetylated microtubule with ratio $\frac{k^+_0}{k^-_0}=221$ \cite{Nishiyama2002} and $\frac{k^+_0}{k^-_0}=802$ \cite{Carter2005} respectively. The error in $\gamma_0$, obtained by simultaneously fitting $P(n)$ and $P(v)$ (Fig. 1), is estimated using the bootstrap sampling method \cite{bookbootstrap}. (c)-(g) are rates at different values of load, calculated using the $F=0$ values in (a).}
\begin{tabular}{lccccc}
  & F & $k^+$  & $k^-$ & $\gamma$\\ [-0.25 em]
  & (pN) & $(s^{-1})$ & $(s^{-1})$ & $(s^{-1})$ \\
 \hline
(a) & 0   & 133.4   & 0.6   & 2.3$\pm$0.3   \\
(b) & 0   & 133.0   & 0.2   & 2.3$\pm$0.3   \\
(c) & 3 & 47.1 & 1.8 & 6.2	\\
(d) & 4 & 33.3 & 2.5 & 8.7   \\
(e) & 5 & 23.5 & 3.6 & 12.1	\\
(f) & 7.6 & 9.5 & 9.3 & 29.0\\
(g) & 8.5 & 7.0 & 12.8 & 39.1 \\
\end{tabular}
\label{tablerates}
\end{table}

We first analyze the $F=0$ data for kinesin-1 (Kin-1)
\cite{Fischermeier2012} using our theory. The motility of Kin-1,
belonging to a family of motors that walks on microtubule filaments, has
been extensively studied since its discovery
\cite{Vale1985}. Experiments have shown that Kin-1 walks
hand-over-hand \cite{Yildiz2004a,Asbury2003} taking discrete steps in
multiples of $s = 8.2 nm$ with each step being almost
identical \cite{Svoboda1993,Kojima1997,Coy1999,Fehr2008}.  The 8.2~nm
is commensurate with the $\alpha/\beta$ tubulin periodicity of a
single MT protofilament, here modeled as a one dimensional lattice
(Fig \ref{fig1}A) \cite{Ray1993}. For Kin-1 the measured $F=0$ mean
velocity is 1089~nm/s (Fig. 2B), which implies that
$k^+_0-k^-_0=132.8$~step/s.  The ratio $\frac{k^+_0}{k^-_0}$ is not
reported in \cite{Fischermeier2012}, which forced us to use data from
other sources. Solving these two equations, we get $k^+_0$ and $k^-_0$
corresponding to two experimental values for $\frac{k^+_0}{k^-_0}=221$
\cite{Nishiyama2002} or $\frac{k^+_0}{k^-_0}=802$ \cite{Carter2005}
(Table I (a) and (b), note that even though the ratios are different,
$k^+_0$ and $k^-_0$ are similar.) Using $k_0^{+}$ and $k_0^{-}$, we
obtained the $F=0$ detachment rate $\gamma_0$ by simultaneously
fitting the measured \cite{Fischermeier2012} velocity distribution
using Eq.~\ref{2Pv>} and run-length distribution using Eq.~\ref{2Pn}.
The excellent fits to both data sets (Fig.~\ref{figwalterdata}) using
a {\it single} parameter shows that our theory captures the basic
aspects of Kin-1 motion. More importantly, the value of the only
unknown parameter $\gamma_0$ is the same in Table~\ref{tablerates}(a)
and (b) and in very good agreement with an independent way of
obtaining the detachment rate (see SI). While analyzing the
experimental data, we convert velocity in step/s to nm/s (or vice
versa) by multiplying (dividing) by $s=8.2$~nm.

The distribution $P(v)$ has the same form as in Eq. \ref{2Pv>} when
the motor is subjected to an external force, $F$, (Fig. \ref{fig1}C)
except $k^{+}(F)$ and $k^{-}(F)$ are dependent on $F$. We model these
rates using the Bell model,
$k^\pm (F)=k^\pm _0 e^{-\frac{F_\parallel d^\pm _\parallel}{kT}}$
where $F_{\parallel}$ is the component of the force parallel to the
microtubule, $k^+_0$ and $k^-_0$ are the forward and backward rates at
$F=0$, and the transition state distances $d^+_\parallel$ and
$d^-_\parallel$ are defined in Fig. \ref{fig1}D.  We can rewrite the
arguments of the exponentials as
$k^\pm(F)=k^\pm_0 e^{-\frac{F d^\pm}{kT}}$, defining effective
distances $d^\pm = d^\pm_\parallel |F_\parallel/F|$.

We obtained $|d^+|=1.4$ nm and $|d^-|=1.6$ nm (Fig~S5), by fitting the
average velocity as a function of force \cite{Nishiyama2002} with
$\overline{v}(F)=(k^+(F)-k^-(F))$ subject to the constraint
$|d^+|+|d^-|=2.9$~nm \cite{Nishiyama2002}, assuming that $d^\pm$ are
independent of $F$. Note that $|d^+|+|d^-|=2.9$~nm is different from
the mean step-size ($8.2$~nm) of Kin-1, because the force transmitted
to the kinesin heads is not parallel to the direction of the motor
movement (Fig~\ref{fig1}C), so
$d^\pm = d^\pm_{\parallel} |F_{\parallel}/F| < d^\pm_{\parallel}$. An
alternate mechanism for $|d^+|+|d^-|$ being less than $8.2$~nm has
been proposed elsewhere \cite{Hyeon09PCCP}. Similarly, the $F$-dependent
detachment rate is taken to be $\gamma(F)=\gamma_0\exp\left(\frac{F_{\perp} d_{\gamma}}{kT}\right)$,
which we write as 
$\gamma(F)=\gamma_0\exp\left(\frac{|F|}{F_d}\right)$ where $F_d = \frac{|F|kT}{F_{\perp}d_{\gamma}}$ ($\approx$ 3 pN)  is
the force at which the two-headed kinesin disengages from the
microtubule
\cite{bookdetachmentrate}. At distances greater than the transition state distance $d_{\gamma}$ the motor is unbound from the MT.  Table \ref{tablerates} (c)--(g), listing
the three rates at several values of $F$, shows that $\gamma(F)$ is
appreciable relative to $k^{+}(F)$ at $F > 3$~pN, which has profound
consequences on $P(v)$, as we show below.

\begin{figure}[t]
\centerline{\includegraphics[width=\columnwidth]{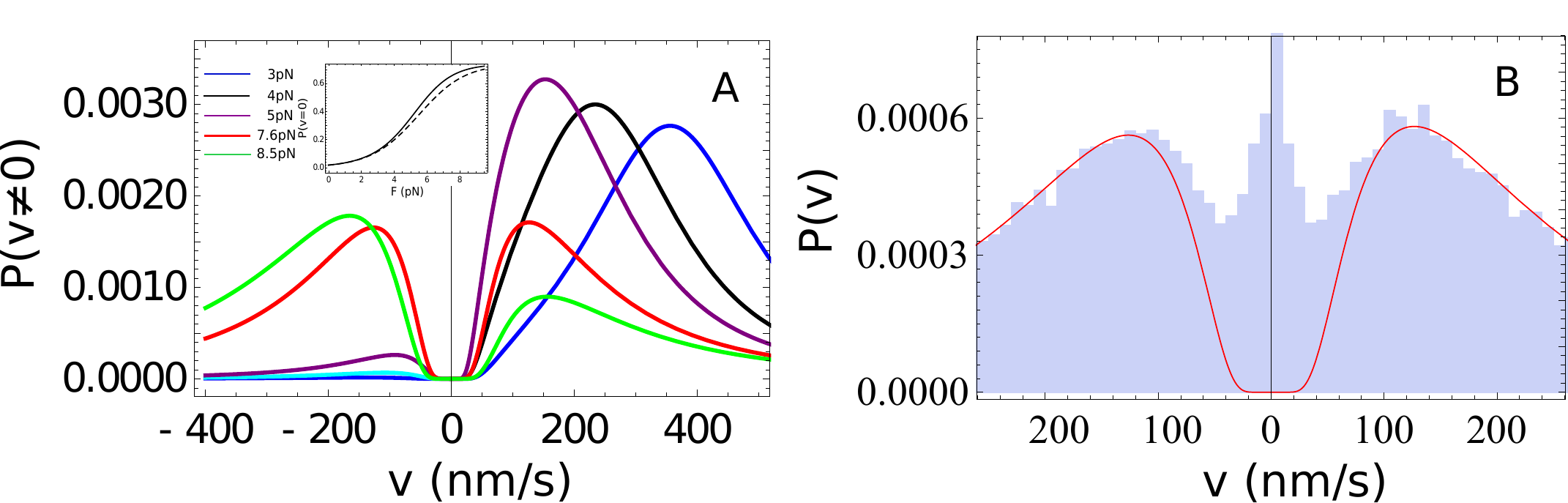}}
\caption{{\bf A:} Predictions of the normalized velocity distributions
  ($(P(v>0) + P(v<0))/(1 - P(v=0))$) at different force values. The
  inset shows $P(v=0)$ as a function of $F$. The dashed line is the
  probability ($\gamma/k_T$) that the motor detaches without taking a
  step.  The solid line is the cumulative probability,
  $P(n=0)=P(v=0)=\frac{\gamma}{\sqrt{k_T^2-4k^+k^-}}$, the total
  number of motors detaching with zero average velocity and zero net
  displacement, which includes both motors that detach without
  stepping, and those that return to their starting location. {\bf B:}
  Kinetic Monte Carlo simulations (blue histograms) of the velocity
  distribution at 7.6 pN load, with a Gaussian step-size distribution,
  experimentally determined for Kin-1 to have a mean of 8.2 nm and
  standard deviation of 1.6 nm \cite{Fehr2008}. The solid lines in both
  graphs are the exact $P(v)$--Eq.~\ref{2Pv>} obtained by assuming that all motors take identical 8.2 nm step. Although inclusion of distribution in the step size leads to minor deviations in $P(v)$ the predicted bimodality persists.}
\label{figpvforce}
\end{figure}

The normalized $P(v\neq 0)$ distributions for different $F$ values are
plotted in Fig.~\ref{figpvforce}A using Eq.~\ref{2Pv>}, showing
distinctly non-Gaussian behavior, in sharp contrast to the
approximately Gaussian distribution at $F=0$ (Fig. 2B). By stringent
standards even at $F=0$, $P(v)$ is not a Gaussian \cite{Hughes2013}
but the extent of deviation from a Gaussian increases dramatically as
$F$ increases. This happens because $\gamma$ increases and eventually
becomes larger than the other rates (Table \ref{tablerates}), thus
decreasing kinesin's processivity (Fig. S5-D). As a result, CLT does not hold resulting in $P(v)$ to exhibit non-Gaussian behavior.  

More unexpectedly, the predicted $F$-dependent $P(v)$s are bimodal
(Fig.~\ref{figpvforce}A). As $F$ increases, the peak at $v<0$ becomes
higher and reaches the same height as the one at $v>0$ at the stall
force $F_S$. For forces below $F_S=7.63$~pN \cite{Nishiyama2002}, the
location of the peak of the $P(v>0)$ curves, $v_P$, shifts to lower
velocity values as $F$ increases, but then moves to higher ones at
$F>F_S$ ($v_P$ at $F=8.5$~pN (green curve) is larger than $v_P$ at
$F_S=7.6$~pN (red curve)).

These two counter-intuitive results are direct consequences of the
discrete step size of kinesin on the MT. For large values of
$\gamma(F)$, corresponding to large forces (where the bimodal
structure is most prominent), the time the motor spends on the
microtubule is necessarily small ($t \sim 1/\gamma(F)$). Since $n s$
has to be an integer multiple of 8.2 nm, it cannot be less than 8.2
nm, implying that velocities close to zero (both positive and
negative) are improbable, giving us a full explanation of the two-peak
structure in $P(v)$.  In addition, for $F\le5$ pN, $v_P$ can be
estimated using 8.2$(k^{+}(F) - k^{-}(F))$~nm/s. As $F$ increases,
$k^+(F)$ decreases while $k^-(F)$ increases (Table I, Fig.~1D),
leading to decrease of $v_P$ with $F$.  However, $v_P$ cannot shift to
arbitrarily low values of the velocity due to the discreteness of the
step-size. As the force increases beyond $F_S$, for most of the
trajectories that contribute to the $v>0$ peak, the motor falls off
after taking just one step (smallest $n$) (Fig.~S5D), and at the same
time the detachment time continuously decreases, shifting $v_P$ to
larger velocities ($ v \sim \frac{1}{t}$).

%\subsection{Velocity distribution with finite step-size distribution}

The discrete nature of the stepping kinetics is less significant when
a large number of $n$ terms contribute to $P(v)$.  The results for
$P(n)$ in Fig. S5D show that at $F=0$, Kin-1 takes in excess of 50
steps of net displacement before detaching. It is then reasonable to replace the
summation in Eq.~\ref{2Pv>} by an integral. An {\it ansatz} for the
approximate velocity distribution, $P_A(v)$, which is highly accurate
for $F=0$ where $\gamma \ll k^+$ (Fig.~2B), is given by
\begin{equation}
P_A(v>0)=\frac{\gamma}{v^2}(1+4a)^{-1/4}\left[2\ln \left(\frac{1+\sqrt{1+4a}}{2}\cdot \frac{v}{k^+}\right)-2\sqrt{1+4a}+2\frac{k_T}{v}\right]^{-3/2},
\label{2Pv3}
\end{equation}
where $a=\frac{k^+k^-}{v^2}$. However, for $F\ne0$ the values of
$\gamma$ are such that only few terms in Eq.~\ref{2Pv>} are
non-negligible, thus making the approximate expression
(Eq.~\ref{2Pv3}) invalid (Fig.~S2).  As $F$ increases the average
number of forward steps decreases dramatically. At $F=3$~pN the
probability of Kin-1 taking in net displacement for more than 10 steps is small.  This results
in qualitative differences between $P_A(v)$ (continuum steps) and
$P(v)$ (discrete steps), which is dramatically illustrated by
comparing Figs.~3A and Fig.~S3. Thus, the discreteness of the motor
step must be taken explicitly into account when analyzing data,
especially at non-zero values of the resistive force.

Although we assumed that stepping occurs in integer multiples of
8.2~nm, experiments show that the step-size distribution has a finite
but small width for Kin-1\cite{Fehr2008}, which could possibly affect
the bimodal structure in $P(v)$ at $F\neq 0$. Since inclusion of the
distribution in the step-sizes makes the model analytically
intractable, we performed kinetic Monte Carlo (KMC) simulations
\cite{Bortz75JCP}. In this set of simulations, every time kinesin
jumps forward or backward, we assume that the step-size has a Gaussian
distribution with mean $s = 8.2$ nm, and standard deviation $\sigma$,
varied from 0 nm to 8 nm. As $\sigma$ increases, the two-peak
structure gets washed out. However, for the experimentally measured
value of $\sigma=1.6$ nm, the two-peak velocity distribution is
present (Fig. \ref{figpvforce}B and S4C) (we converted the standard
error $SE=0.03$~nm from the measurements (see Fig.~2 in
\cite{Fehr2008}) on wild type kinesin LpK to
$\sigma=SE\sqrt{N}=0.03\sqrt{2993}=1.6$~nm). We conclude that the key
predictions in Fig. \ref{figpvforce}A are robust with respect to
inclusion of a physically meaningful step-size distribution and
experiments should be able to discern the two-peak structure in $P(v)$
at $F \ne 0$. The predictions for $P(v)$ at various forces
with $\sigma=1.6$nm are displayed in Fig.~S5A--C.
 
To test the effect of intermediate states, which apparently are needed
to analyze experiments \cite{Fisher2001}, on the bimodality of $P(v)$,
we generalized the model to include a chemical intermediate (Fig. S6)
and computed $P(v)$ exactly (see SI for details). We found that the
important bimodal feature is still preserved, thus further
establishing the robustness of our conclusions.

 For $F \ne 0$ the motor would take only a small number of steps because $\gamma (F)$ is an increasing function of $F$, raising the possibility that the predictions in Fig.~3A may not be measurable. To account for potential experimental limitations we calculated the conditional probability $P(v|n > n_0)$ which is obtained by neglecting the first $n_0$ terms in Eq.~\ref{2Pv>}. Fig.~S10 shows that even with this restriction the bimodal distribution persists especially near the stall force where it is prominent.

There are substantial variations in force velocity ($F$-$v$) curves
for Kin-1 in different experiments. Even the shape of the $F$-$v$
curve in \cite{Nishiyama2002} does not agree with the results in
\cite{Visscher1999,Carter2005}. For completeness, we analyzed the
$F$-$v$ curves reported in \cite{Carter2005,Visscher1999}. The
theoretical fits are in excellent agreement with experiments, which
allows us to calculate $P(v)$ at $F\ne0$. The results in Fig.~S7 show
that, just as those in Fig.~3A, the predicted bimodality in $P(v)$
remains regardless of the differences in the shapes of the $F$-$v$
curves among different experiments.

In summary, exact theoretical analysis using a simple model for motor
motility quantitatively explains the zero force velocity and
run-length distributions {\it simultaneously} for Kin-1, with just one
physically reasonable fitting parameter. With an average run-length of
$\sim$632~nm = 77~steps \cite{Fischermeier2012}, we expect the
velocity distribution of Kin-1 to deviate from a Gaussian (albeit
slightly) even at zero force \cite{Hughes2013}. Based on the analysis
of the zero force data, we calculated the load dependence of the
velocity distribution of Kin-1 and discovered that the discrete nature
of kinesin's steps leads to an unexpected bimodal structure in the
velocity distribution under load. This surprising result can be tested
in single molecule experiments, most readily accessed near the stall
force where the motor has equal probability of moving forward or
backward. An example of such a trajectory may be found in Fig.~1 of
\cite{Carter2005}. It remains to be seen if our predictions can be
readily tested within the precision of single molecule experiments.
Although set in the context of Kin-1, our general theory can be used
to analyze experimental data for any molecular motor for which the
model in Fig. \ref{fig1}B is deemed appropriate.  As a result,
our major results should hold for any finitely processive motor that
takes discrete steps.

\clearpage

\section{Supplementary Information} 

{\bf Distribution of run-length, $P(n)$:}
The equation for run-length distribution $n>0$ (Eq. 2 in the main text) can be written as, 
\begin{eqnarray}
P(n>0)&=&\left(\frac{k^+}{k_T}\right)^{n}\left(\frac{\gamma}{k_T}\right)\sum_{l=0}^{\infty}\left(\frac{k^+k^-}{k_T^2}\right)^l\frac{(2l+n)!}{(n+l)! l!}\\
&=&\left(\frac{k^+}{k_T}\right)^{n}\left(\frac{\gamma}{k_T}\right) {}_2\textbf{F}_1(\frac{1+n}{2},\frac{2+n}{2};1+n;4\frac{k^+k^-}{k_T^2}).
\end{eqnarray}
In the above equation ${}_2\textbf{F}_1$ is a Gaussian hypergeometric function. Using the special case of Gaussian series for ${}_2\textbf{F}_1$ (page 556 of \cite{abramowitz+stegun}) leads to  Eq. (3) of the main text.

{\bf Derivation of the velocity distribution ($P(v))$:}
Consider a trajectory in which a motor takes $m$ forward and $l$ backward  steps  at times $t_1, t_2, ..., t_{m+l}$ respectively, before detaching at $t$. The times are ordered such that  $t_1<t_2<...<t_{m+l}<t$. There are $\frac{(m+l)!}{m!l!}$ combinations in which these steps can occur resulting in a net displacement, $n = (m - l)$, of the motor.  The waiting time between steps is much greater than the duration of a step. Consequently, stepping from any site is entirely independent of all other steps but only depends on the total rate, $k_T$, which is the sum of the detachment rate and the rates of forward and backward steps.   Thus, the probability $P(m,l,t_1,t_2,...,t_{m+l},t)$ of realizing such a trajectory is,
\begin{eqnarray}
P(m,l,t_1,\ldots,t_{m+l},t)&=\frac{(m+l)!}{m!l!}&\prod_{i=1}^{m}{\bigg[k_Te^{-k_T(t_i-t_{i-1})}\frac{k^+}{k_T}\bigg]}\cdot\prod_{j=m+1}^{l+m}{\bigg[k_Te^{-k_T(t_{j}-t_{j-1})}\frac{k^-}{k_T}\bigg]}\nonumber \\
&&\cdot\left[k_Te^{-k_T(t-t_{m+l})}\frac{\gamma}{k_T}\right]\\
&=&\frac{(m+l)!}{m!l!}(k_T)^{m+l+1}e^{-k_Tt}\left(\frac{k^+}{k_T}\right)^m\left(\frac{k^-}{k_T}\right)^l\frac{\gamma}{k_T}
\label{PmltSI}
\end{eqnarray}
with $t_0=0$ and $t_{m+l+1}=t$.
The joint probability density for the motor to jump $m$ steps forward, $l$ steps backward before detaching at $t$ is 
\begin{equation}
P(m,l,t)=\int_0^t  dt_1 \int_{t_1}^t dt_2 \ldots \int_{t_{n-1}}^t dt_{m+l} P(m,l,t_1,t_2,...,t_{m+l},t).
\end{equation}
 Upon evaluating the integrals we obtain,
\begin{equation}
%P(m,l,t)&=&\int_0^t P(m,l,t_1,t_2,...,t_{m+l},t)dt_1 \int_{t_1}^t dt_2 \ldots \int_{t_{n-1}}^t dt_{m+l}\\
%&=&(k_T)^{m+l+1}e^{-k_Tt}\left(\frac{k^+}{k_T}\right)^m\left(\frac{k^-}{k_T}\right)^l\frac{\gamma}{k_T}\frac{(m+l)!}{m!l!}\frac{t^{m+l}}{(m+l)!}\\
P(m,l,t)=\left(\frac{t^{m+l}}{m!l!}\right)(k^+)^m(k^-)^l\gamma e^{-k_Tt},
\label{2Pmlt2}
\end{equation}
which is Eq.~4 in the main text. From Eq.~\ref{2Pmlt2} an expression for $P(n,t)$ may be derived by setting $n = m-l$ and performing the sum over $m$ and $l$ resulting in 
\begin{equation}
P(n,t) = \left(\frac{k^+}{k^-}\right)^{\frac{n}{2}}\gamma\exp(-k_Tt)I_{n}(2t\sqrt{k^+k^-})
\label{Pn}
\end{equation}
where $I_n(x)$ is the modified Bessel function of the first kind.

Substituting Eq.~\ref{Pn} into $P(v)=\sum_{n=-\infty}^\infty \int_{0}^\infty
dt\,\delta(v-n/t)P(n,t)$, we get 
\begin{equation}
P(v>0)=\sum_{n=0}^\infty \frac{n}{v^2}\left(\frac{k^+}{k^-}\right)^{\frac{n}{2}}\gamma\exp(-k_T\frac{n}{v})I_{n}(\frac{2n}{v}\sqrt{k^+k^-})\\
\label{2PvI}
\end{equation}
%$B=\frac{\gamma}{v^2}$ and $C=\sqrt{\frac{k^+}{k^-}}\textbf{e}^{-\frac{k_T}{v}}$.
By expressing $I_{n}(\frac{2n}{v}\sqrt{k^+k^-})$ in terms of the hypergeometric function ${}_0F_1$, we obtain (Eq.~5 in the main text),
\begin{equation}
P(v>0)=\frac{\gamma}{v}\sum_{n=0}^{\infty}\left(\frac{n}{v}\right)^{n+1}\frac{1}{ n!}\left(k^+ e^{-\frac{k_T}{v}}\right)^n{}_0F_1\left(;n+1;\frac{n^2k^+k^-}{v^2}\right),
\label{2Pv>2}
\end{equation}
where the hypergeometric function is 
${}_0F_1\left(;n+1;\frac{n^2k^+k^-}{v^2}\right)=\sum_{l=0}^\infty{\frac{1}{(n+1)^{l}}\frac{\left(\frac{n^2k^+k^-}{v^2}\right)^l}{l!}}$.
Following the steps leading to Eq.~\ref{2Pv>2}, the expression for $P(v<0)$ becomes,
\begin{eqnarray}
P(v<0)
&=&\frac{\gamma}{-v}\sum_{n=0}^{\infty}\left(\frac{n}{-v}\right)^{n+1}\frac{1}{ n!}\left(k^- e^{\frac{k_T}{v}}\right)^n{}_0F_1\left(;n+1;\frac{n^2k^+k^-}{v^2}\right)
\label{2Pv<}
\end{eqnarray}

%{\bf Estimatation of the Kin1 detachment rate, $\gamma_0$:}
\begin{figure}
\centerline{\includegraphics[width=0.6\columnwidth]{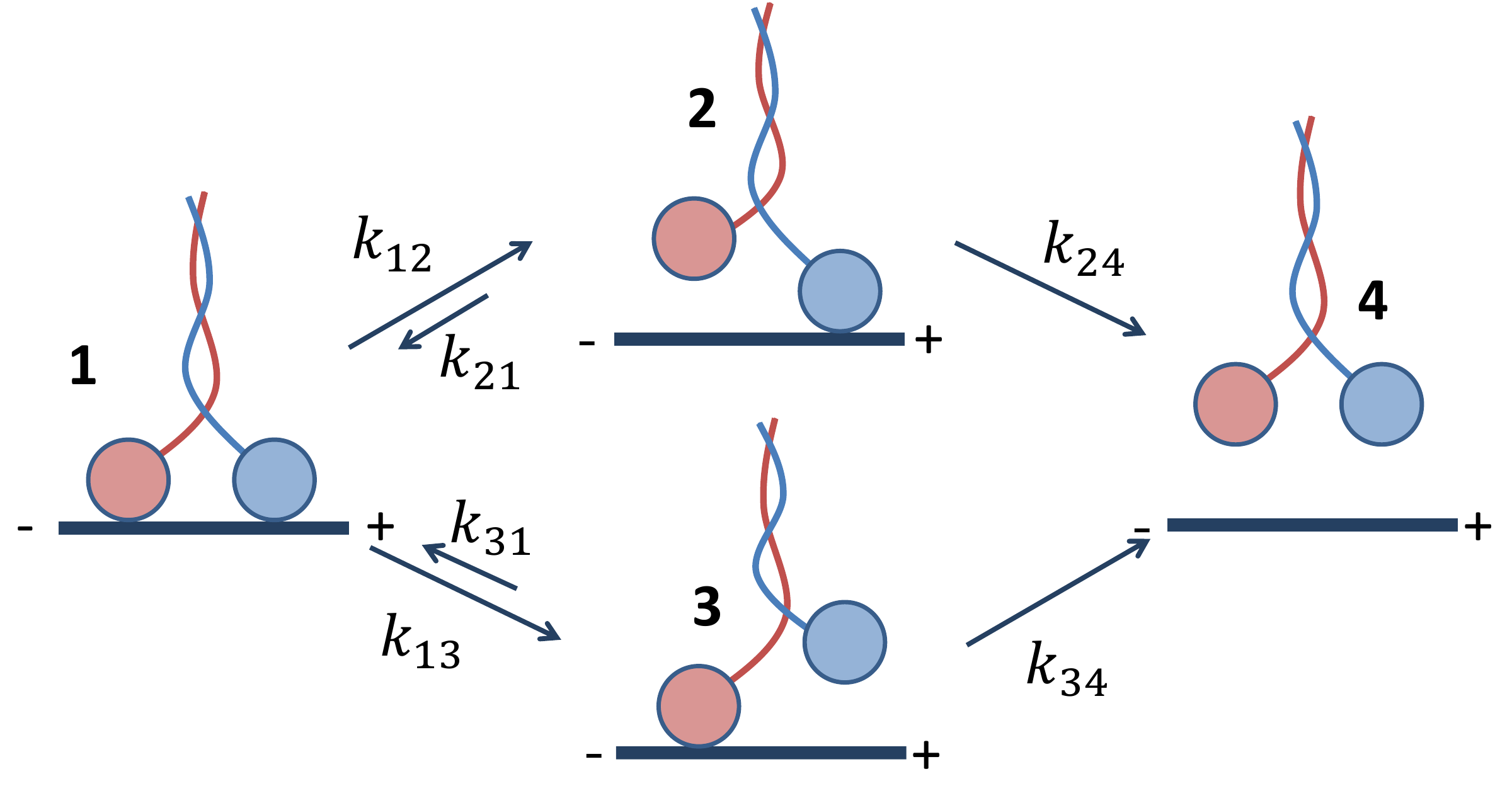}}
\caption{Enzymatic pathway used to estimate the Kin1 detachment rate, $\gamma_0$.}
\label{figdetachmentrate}
\end{figure}
An approximate expression for $P(v)$, which is accurate only for $F=0$ can be obtained by turning the summation into an integral (continuum step size). By using, ${}_0\text{F}_1(;n+1;an^2)\approx(1+4a)^{-1/4}\exp(nc)$ where $c=\sqrt{1+4a}-1-\ln \left(\frac{1+\sqrt{1+4a}}{2}\right)$, and and $a=\frac{k^+k^-}{v^2}$.
%\begin{equation}
%${}_0\text{F}_1(;n+1;an^2)\approx(1+4a)^{-1/4}\exp(nc)$
%\label{hyperaprox}
%\end{equation}
%where
%\begin{equation}
%c=\sqrt{1+4a}-1-\ln \left(\frac{1+\sqrt{1+4a}}{2}\right)
%\end{equation}
%and $a=\frac{k^+k^-}{v^2}$. If $k^-=0$, we can obtain a simple expression for the velocity distribution in the limit $\gamma << k^+$. 
Using these simplifications and the second order Stirling's approximation we obtain,
\begin{equation}
P(v>0)=\frac{\gamma}{v^2\sqrt{2\pi}}(1+4a)^{-1/4}\sum_{n=0}^{\infty}\sqrt{n}A^n
\label{2Pv2}
\end{equation}
where the argument $A=\frac{k^+}{v}\exp\left(\sqrt{1+4a}-\ln\left(\frac{1+\sqrt{1+4a}}{2}\right)-\frac{k_T}{v}\right).$
Converting the sum in Eq.~\ref{2Pv2} into an integral% (see Eq.~\ref{intapprox})
, we arrived at an {\it ansatz} for $P_A(v>0)$, quoted in Eq.~6 of the main text. Similarly,
\begin{equation}
P_A(v<0)=\frac{\gamma}{v^2}(1+4a)^{-1/4}\left[2\ln \left(\frac{1+\sqrt{1+4a}}{2}\cdot \frac{|v|}{k^-}\right)-2\sqrt{1+4a}+2\frac{k_T}{|v|}\right]^{-3/2}
\label{2PAv}
\end{equation}
%Comparison between the analytical results with Kinetic Monte Carlo simulations \cite{KMC2007} in Fig \ref{figsimulation} validates the approximations leading to Eq.~6.

\begin{figure}
\centerline{\includegraphics[width=.6\columnwidth]{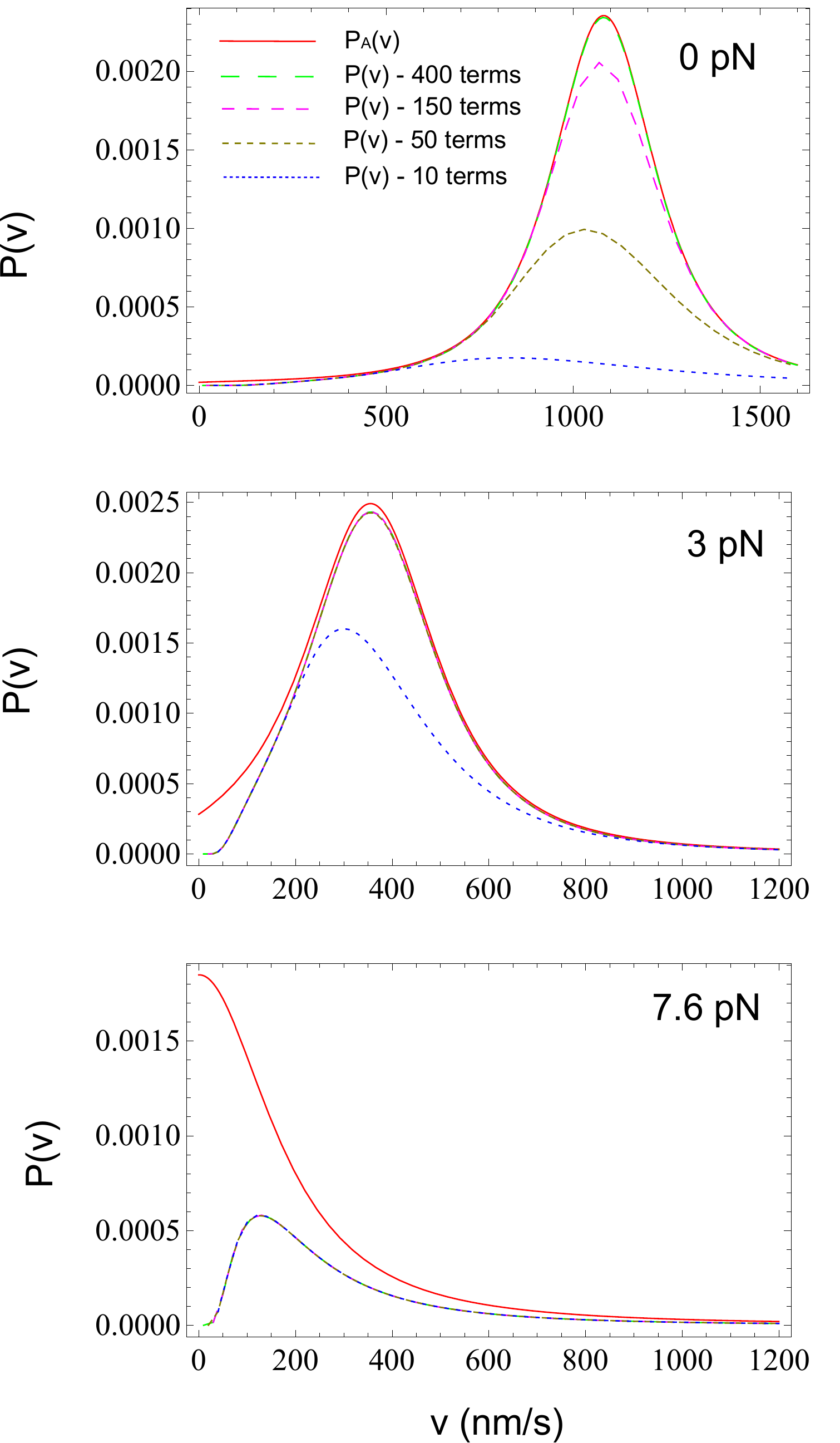}}
\caption{Comparison of the exact (Eq.~\ref{2Pv>2} plotted as dashed curves) and approximate (Eq.~\ref{2PAv} plotted in red) velocity distribution for three  values of $F$. The number of terms needed to converge the sum in Eq~\ref{2Pv>2} is $F$-dependent. As a test of convergence we show $P(v)$ as a function of the number terms used in the summation in Eq.~\ref{2Pv>2}. The maximum sum used in the obtaining the dashed curves are shown in the top panel.  For $F=0$ in excess of 400 terms are needed whereas at higher $F$ converged results are obtained using very few terms. Comparison of the results for $P(v)$ and $P_A(v)$ shows agreement only for $F=0$. There are qualitative differences between $P(v)$ and $P_A(v)$ as $F$ increases.}
\label{figpavpv}
\end{figure}

{\bf Estimation of the Kin1 detachment rate, $\gamma_0$:}
To estimate the detachment rate of kinesin, we consider the pathway  in Fig \ref{figdetachmentrate}, where kinesin from a State 1 with both heads bound to the microtubule (MT), can detach one head at a time (States 2 and 3), and before fully dissociating from the track. The detachment rate of the two-headed kinesin is $k_{total}=k_{14}= \gamma$ that can be estimated from the microscopic rates $k_{ij}$, defined in Fig.~\ref{figdetachmentrate}. We assume that once the two heads dissociate, the rates for reattachment to the MT are zero ($k_{43}=k_{42}=0$). The Laplace transform $\tilde{\pi}_i(s)$ of $\pi_i(t)$, which is probability density distribution that the system goes from state $i$ to the final dissociated state 4 in time $t$ \cite{Shaevitz2005}, is
\begin{eqnarray}
\tilde{\pi}_2(s)&=&\frac{k_{21}\tilde{\pi}_1(s)+k_{24}\tilde{\pi}_4(s)}{k_{21}+k_{24}+s}\\
\tilde{\pi}_3(s)&=&\frac{k_{31}\tilde{\pi}_1(s)+k_{34}\tilde{\pi}_4(s)}{k_{31}+k_{34}+s}\\
\tilde{\pi}_1(s)&=&\frac{k_{12}\tilde{\pi}_2(s)+k_{13}\tilde{\pi}_3(s)}{k_{12}+k_{13}+s}\\
\tilde{\pi}_4(s)&=&1.
\end{eqnarray}
Solving for the distribution $\tilde{\pi}_{total}(s)=\tilde{\pi}_1(s)$ yields
\begin{equation}
\tilde{\pi}_{total}(s)=\frac{k_{13} k_{34} (k_{21} + k_{24} + s) + k_{12} k_{24} (k_{31} + k_{34} + s)}{k_{12} (k_{24} + s) (k_{31} + k_{34} + s) + (k_{21} + k_{24} + s) (k_{13} (k_{34} + s) + 
    s (k_{31} + k_{34} + s))}.
\end{equation}
The average detachment time $<t>=\gamma^{-1}$ is the first moment of the generating function $\tilde{\pi}_{total}(s)$,
\begin{equation}
<t>=\frac{1}{\gamma}=\left(-\frac{d\tilde{\pi}_{total}(s)}{ds}\right)\bigg\vert_{s=0}=\frac{k_{13} (k_{21} + k_{24}) + (k_{12} + k_{21} + k_{24}) (k_{31} + k_{34})}{
k_{13} (k_{21} + k_{24}) k_{34} + k_{12} k_{24} (k_{31} + k_{34})}
\end{equation}
In the limit $k_{21}\rightarrow0$ and $k_{31}\rightarrow 0$, which would result in an over estimate of $k_{14}$, we obtain,
\begin{equation}
\gamma<k_{14}=\frac{
k_{13} k_{24} k_{34} + k_{12} k_{24}  k_{34}}{k_{13}  k_{24} + (k_{12} + k_{24})  k_{34}}.
\end{equation}
The microscopic rates $k_{ij}$ for the Kin1 have been reported in an experimental paper \cite{Hancock1999}. We expect the detachment rate of the kinesin trailing head (State 2) to be faster so we set $k_{12}=48s^{-1}$ from their measurements. All other rates are set to the one-headed detachment rate $k_{13}=k_{24}=k_{34}=3s^{-1}$. With these values, the estimated detachment rate of Kin1 is $\gamma_0=2.8s^{-1}$. This value is in a very good agreement with $\gamma_0$ ($=2.3s^{-1}$) extracted by using  our theory to simultaneously fit the  experimental data for $P(v)$ and $P(n)$.

{\bf }
\begin{figure}
\centerline{\includegraphics[width=0.6\columnwidth]{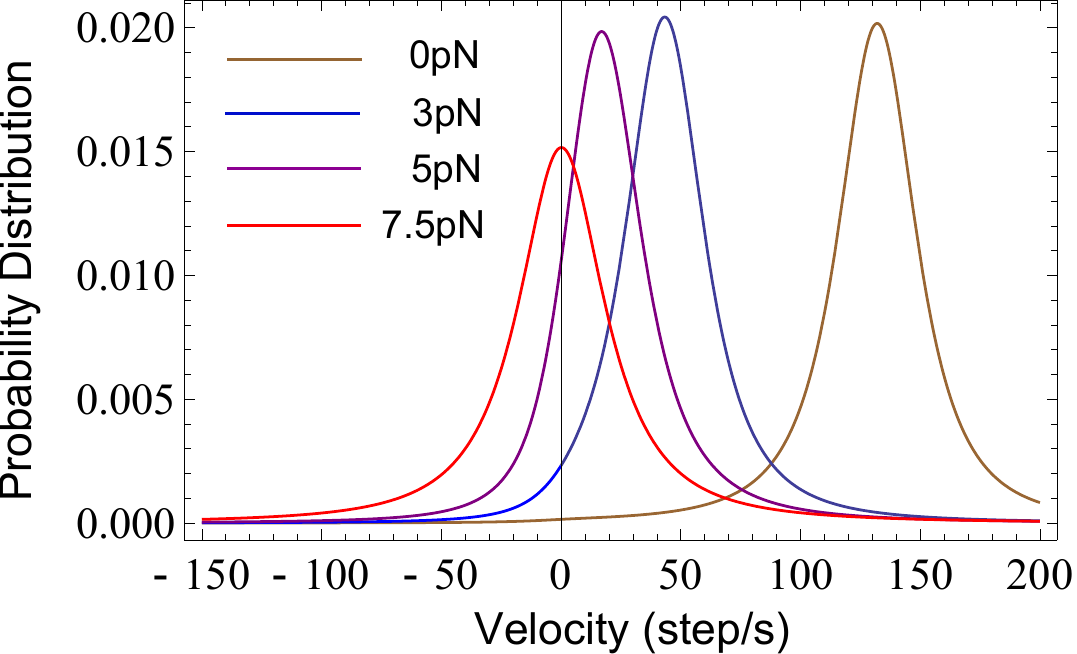}}
\caption{Predictions of velocity distributions at different values of loads using the approximation Eq.~\ref{2PAv}. The continuum approximation washes out the bimodal structure in the exact $P(v)$ (see Fig.~3 in the main text).}
\label{figintegral}
\end{figure}

\begin{figure}[h]
\centerline{\includegraphics[width=1\columnwidth]{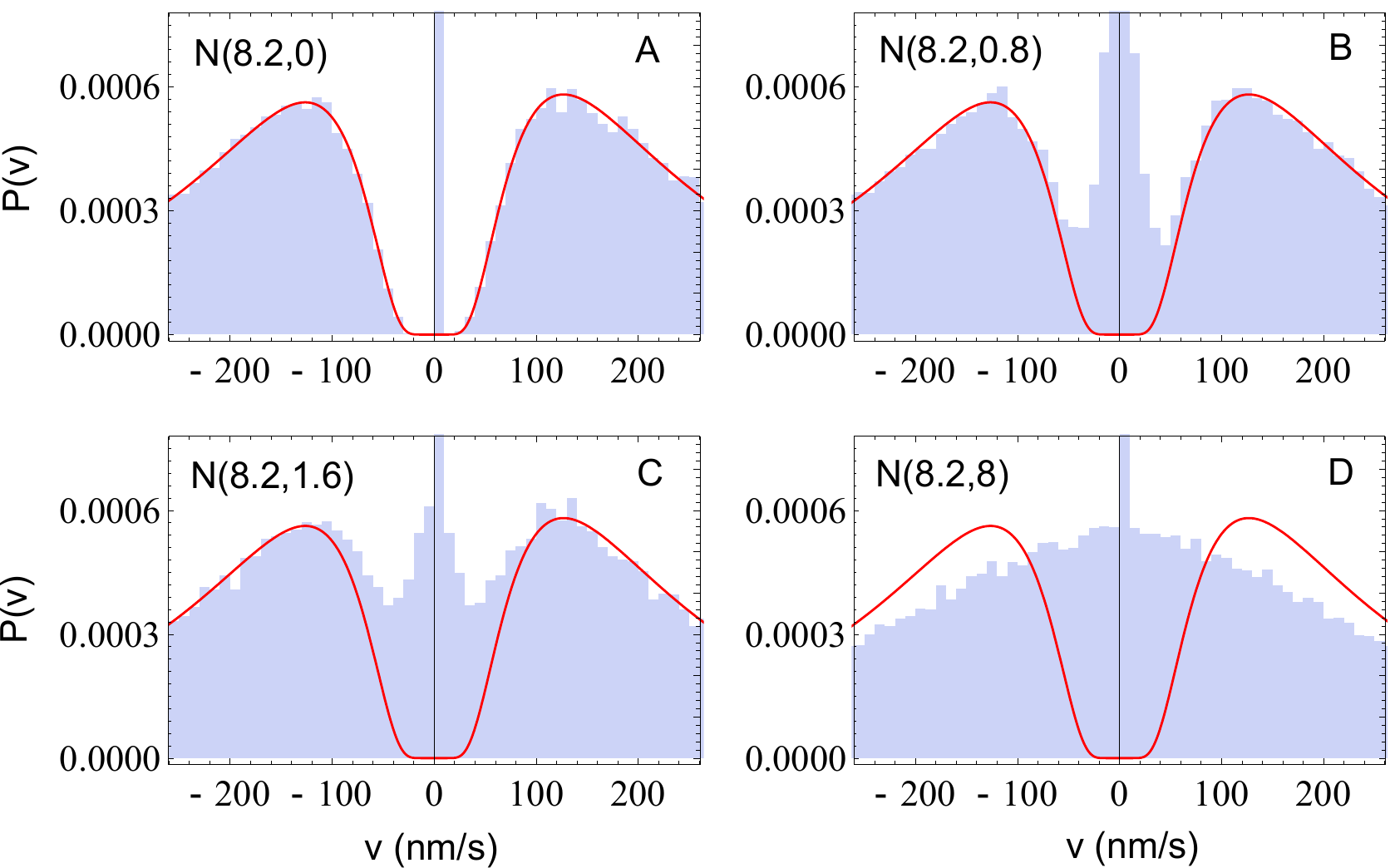}}
\caption{Progressive loss of bimodal nature of the velocity distribution at 7.6~pN load, with increasing width of the step-size distribution. N(m,$\sigma$) in the left corner of each panel denotes the Gaussian distribution with mean $s$ and standard deviation $\sigma$. The solid line is the exact velocity distribution function (Eq.~11) from our model, with no step-size distribution. The blue histograms are computed using Kinetic Monte Carlo simulations.}
\label{figstepsize}
\end{figure}

\begin{figure}[h]
\centerline{\includegraphics[width=1\textwidth]{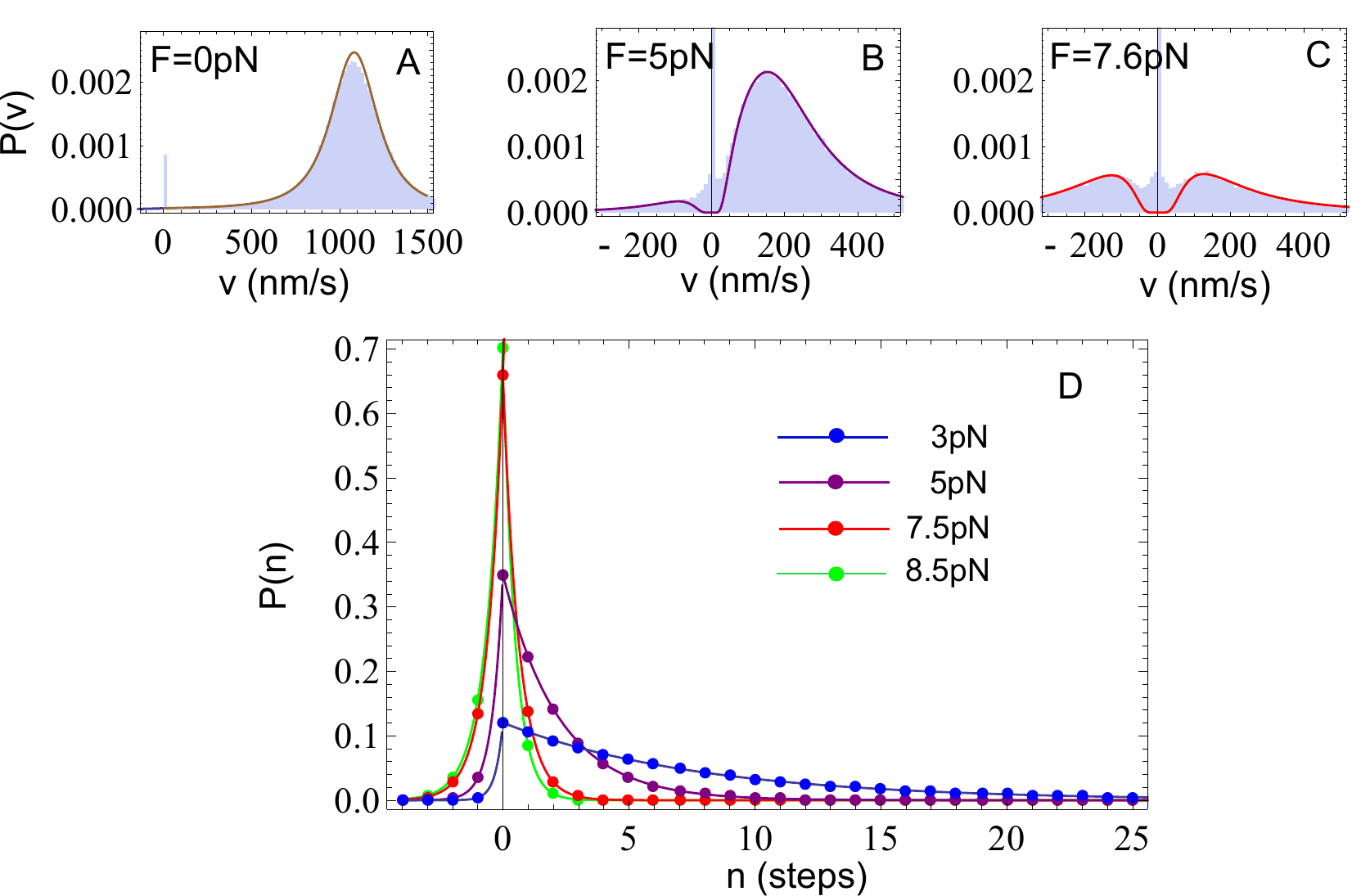}}
\caption{{\bf A--C:} Velocity distribution simulations with step-size distribution N(8.2,1.6) with increasing load (blue histograms). The solid lines, which are the same exact velocity distribution functions (Eq.~11) with no step-size distribution, fit the simulation results quantitatively. {\bf D:} Predictions of runlength distributions at different values of load.}
\label{figpvstepforce}
\end{figure}

{\bf Kinetic Monte Carlo (KMC) Simulations:}
In order to account for the possible distribution in step sizes noted in experiments we performed KMC simulations to calculate the $P(v)$.  In these simulations, we assume that stepping and detachment  first-order are processes with exponential decay statistics. The decision to execute a forward or backward step or disassociate from the track is made by sampling time from $P(\Delta t)=k_T e^{-k_T \Delta t}$. If the motor  takes a step, its  $s$ is sampled from the Gaussian distribution $N(s,\sigma)$, and the waiting time at the new binding site for $\Delta t$  is drawn from $P(\Delta t)$.  A single stochastic trajectory corresponds to the movement of the motor before detaching.  The velocity for a trajectory $j$ is calculated using $v_j=\sum_i s_i/\sum_i \Delta t_i$). The distribution $P(v)$ is obtained by computing $v_j$ for an ensemble of trajectories (see Fig.~\ref{figstepsize} for $P(v)$ at $F_s$ as a function of $\sigma$ the dispersion in $N(s,\sigma)$).
Results for the velocity distribution functions at two other values of $F$ along with the simulation results using $N(8.2,1.6)$ are given in Fig.~\ref{figpvstepforce}.

\begin{figure}[h]
\centerline{\includegraphics[width=1\textwidth]{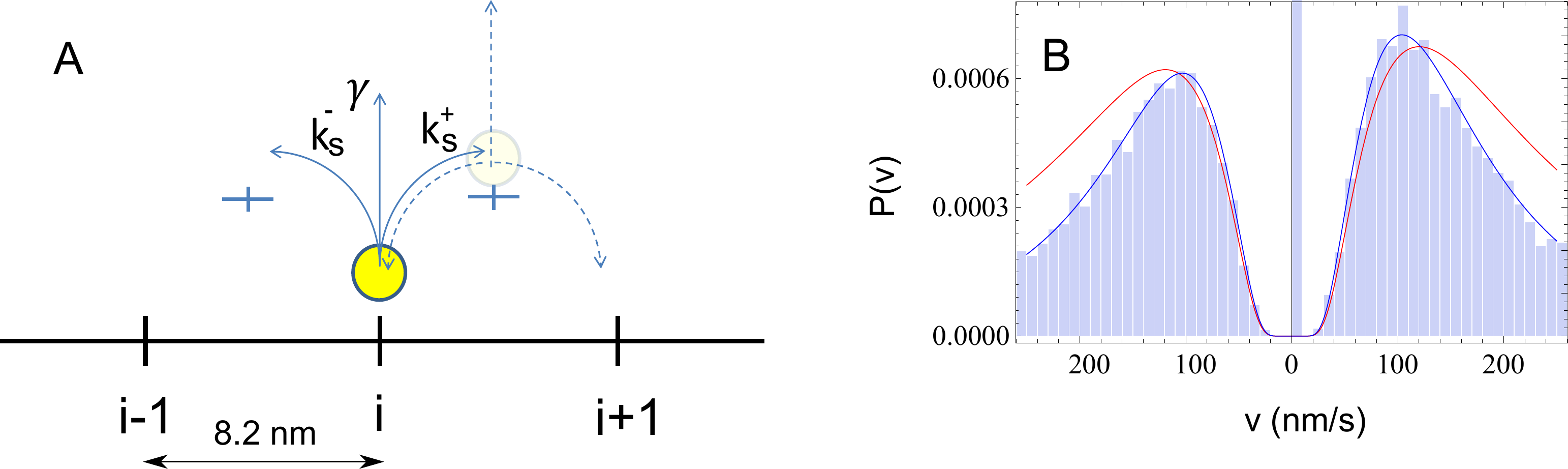}}
\caption{{\bf A:} Kinetic model for stepping with an intermediate state at every step. {\bf B:} Velocity distribution simulations with intermediate state from KMC simulations (blue histograms) and the exact function Eqs.~\ref{Pv>inter} and \ref{Pv<inter} (blue line). For comparison we also show the exact velocity distribution function for the model (Eq.~\ref{2Pv2}) with no step-size distribution (red line). The value of $F=7.5$~pN. Although the velocity distribution becomes narrower when an intermediate state is present, the two-peak structure still remains.}
\label{figinterstate}
\end{figure}

{\bf Velocity distribution with an intermediate state:}
It is natural to wonder if the predicted bimodality in $P(v)$ at $F \ne 0$ is a consequence of the simplicity of the model. We now show that the presence of a chemical intermediate state, which has been invoked to analyze experiments \cite{Fisher2001}, does not alter the bimodal structure in $P(v)$. We consider a model with a chemical  intermediate state (Fig.~\ref{figinterstate}A). Each motor steps from site $i$ to site $i+1$, which is physically $\SI{8.2}{nm}$ away along the track, by visiting a single chemical intermediate state. In any state, the motor can take a sub-step to the next state with the same forward rate $k^+_s$, backward rate $k^-_s$, and detachment rate $\gamma$. 

The probability for the motor to take a single step forward, by passing the intermediate state and subsequently binding to the site $\SI{8.2}{nm}$ away, is $P_i(1,t)=t(k^+_s)^2e^{-k_Tt}$
where $k_T=k^+_s+k^-_s+\gamma$. The probability for the motor to take $m$ step forward, $l$ step backward and then detach from the track at time $t$ is,
\begin{align}
P_i(m,l,t)&=\bigg[\frac{t^{2(m+l)}}{(2m)!(2l)!}(k^+_s)^{2m}(k^-_s)^{2l}
+\frac{t^{2(m+l)+1}}{(2m+1)!(2l)!}(k^+_s)^{2m+1}(k^-_s)^{2l}	\nonumber\\
&+\frac{t^{2(m+l)+1}}{(2m)!(2l+1)!}(k^+_s)^{2m}(k^-_s)^{2l+1}	
+\frac{t^{2(m+l+1)}}{(2m+1)!(2l+1)!}(k^+_s)^{2m+1}(k^-_s)^{2l+1}\bigg]\gamma\exp({-k_Tt})
\end{align}
Thus, the probability for the motor to complete $n=(m-l)$ steps in time $t$ becomes,
\begin{align}
P_i(n,t)&=\sum_{m,l=0}^\infty \delta_{(m-l),n}P_i(m,l,t)	\nonumber\\
=&\gamma e^{-k_T t} t^{2n} (k^+_s) ^{2n} \sum_{l=0}^\infty \bigg[
\frac{t^{4l}(k^+_sk^-_s)^{2l}}{(2n+2l)!(2l)!}
+tk^+_s\frac{t^{4l}(k^+_sk^-_s)^{2l}}{(2n+2l+1)!(2l)!}	\nonumber\\
&+tk^-_s\frac{t^{4l}(k^+_sk^-_s)^{2l}}{(2n+2l)!(2l+1)!}
+t^2k^+_sk^-_s\frac{t^{4l}(k^+_sk^-_s)^{2l}}{(2n+2l+1)!(2l+1)!} \bigg]
\end{align}
The velocity distribution for $P_i(v>0)$ is,
\begin{align}
P_i(v>0)=&\frac{\gamma}{v}\sum_{n=0}^\infty \bigg[\bigg(\frac{n}{v}\bigg)^{2n+1}\big( (k^+_s)^2e^{-\frac{k_T}{v}}\big)^n\frac{1}{2k^+_sk^-_sn}\bigg(-(k^+_s+k^-_s)v{}_0F_1\big[2n,-\frac{k^+_sk^-_sn^2}{v^2}\big]	\nonumber\\
&+(k^+_s+k^-_s)v{}_0\mathbf{F}_1\big[2n,\frac{k^+_sk^-_sn^2}{v^2}\big]
+2k^+_snv{}_0\mathbf{F}_1\big[2n+1,-\frac{k^+_sk^-_sn^2}{v^2}\big]	\nonumber\\
&+2k^+_sn(k^-_s-v){}_0\mathbf{F}_1\big[2n+1,\frac{k^+_sk^-_sn^2}{v^2}\big]\bigg)\bigg],
\label{Pv>inter}
\end{align}
and 
\begin{align}
P_i(v<0)=&\frac{\gamma}{-v}\sum_{n=0}^\infty \bigg[\bigg(\frac{n}{-v}\bigg)^{2n+1}\big( (k^-_s)^2e^{\frac{k_T}{v}}\big)^n\frac{1}{2k^+_sk^-_sn}\bigg((k^+_s+k^-_s)v{}_0\mathbf{F}_1\big[2n,-\frac{k^+_sk^-_sn^2}{v^2}\big]	\nonumber\\
&-(k^+_s+k^-_s)v{}_0\mathbf{F}_1\big[2n,\frac{k^+_sk^-_sn^2}{v^2}\big]
-2k^-_snv{}_0\mathbf{F}_1\big[2n+1,-\frac{k^+_sk^-_sn^2}{v^2}\big]	\nonumber\\
&+2k^-_sn(k^+_s+v){}_0\mathbf{F}_1\big[2n+1,\frac{k^+_sk^-_sn^2}{v^2}\big]\bigg)\bigg].
\label{Pv<inter}
\end{align}

Eqs.~\ref{Pv>inter} and \ref{Pv<inter} (blue lines) are plotted in Fig.~\ref{figinterstate}B to compare with KMC simulations of the same model (blue histograms). The velocity distributions Eq.~\ref{2Pv>2} and Eq.~\ref{2Pv<} (red lines) for the model (Fig.~1C) in the main text are also plotted for comparison.   In the intermediate state model, the motor takes two sub-steps with rates $k^+_s$, $k^-_s$, to complete a full step, which in the model in the Fig.~1 occurs with $k^+$ and $k^-$ rates. For Kin1 at $F=7.5{pN}$ we chose $k^+_s=2k^+$, $k^-_s=2k^-$ and $\gamma=\SI{2.3}{s^{-1}}$ in Eqs.~\ref{Pv>inter} and \ref{Pv<inter}. The presence of the intermediate state merely makes the velocity distribution narrower without altering the bimodality in $P(v)$.

\begin{figure}
\centerline{\includegraphics[width=1\columnwidth]{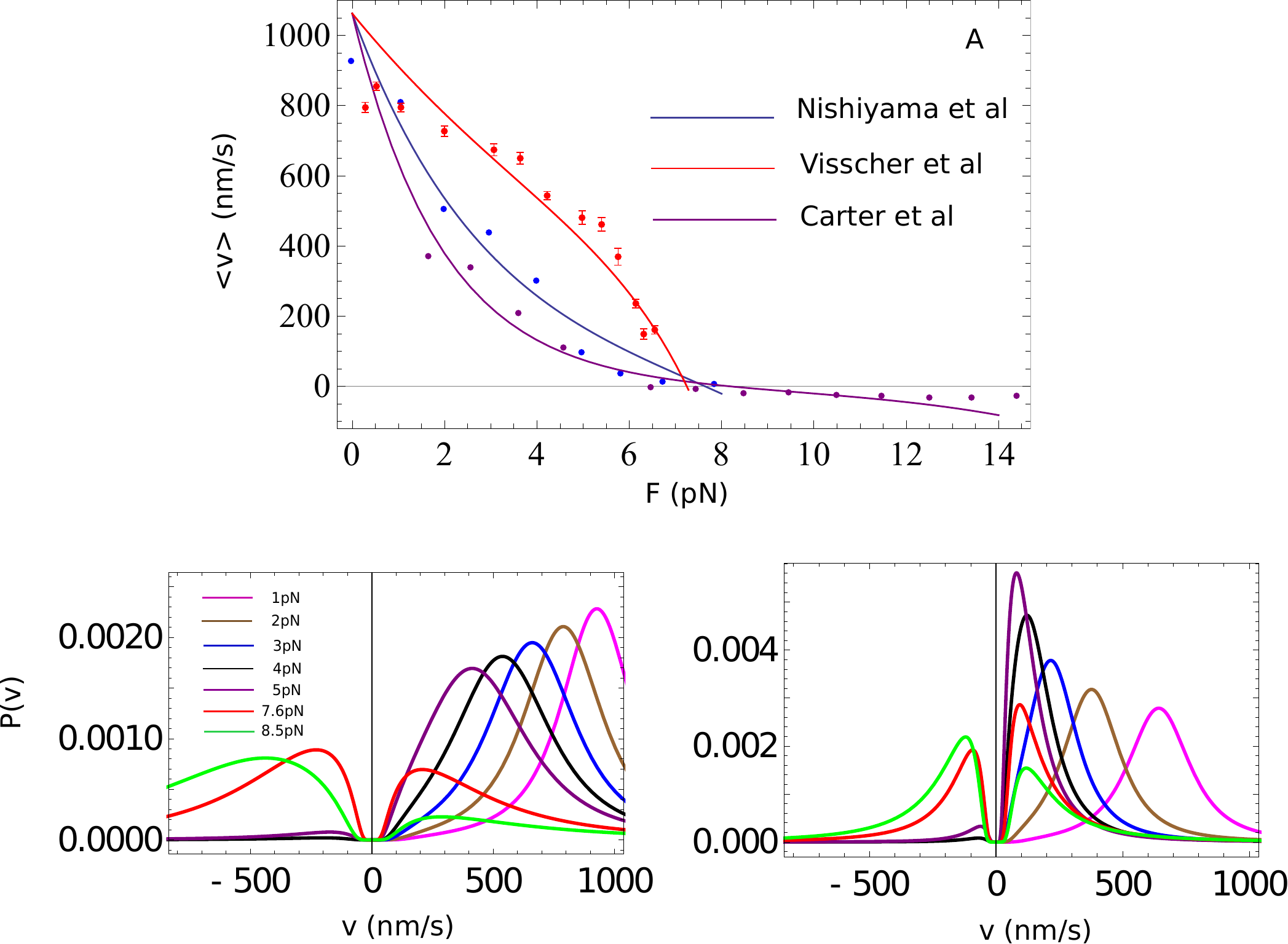}}
\caption{{\bf A:} Force dependence of average velocity of Kin1. The dots are experimental data from \cite{Nishiyama2002} (Blue), \cite{Visscher1999} (Red), \cite{Carter2005} (Purple), while the lines are fits to the  data, using Eq. $\overline{v}(F)=s(k^+_0\textbf{e}^{-\frac{F d^+}{kT}}
-k^-_0\textbf{e}^{-\frac{F d^-}{kT}})$. The characteristic distances extracted from the fit are in the Table~\ref{tableratesT7}. {\bf B:} Predictions for $P(v)$ at different values of loads with parameters listed in the second row of Table~\ref{tableratesT7}. {\bf C} Same as {\bf B} except the third row parameters in Table~\ref{tableratesT7} are used.}
\label{figBlockYanagidaCross}
\end{figure}

{\bf Establishing bimodal characteristics in $P(v)$ using force-velocity curves:} The shapes of the measured force velocity ($F-v$) curves, at a fixed ATP concentration, vary in different experiments \cite{Nishiyama2002,Visscher1999,Carter2005}. In order to assess their impact on our predictions we used the data from \cite{Carter2005,Visscher1999} for $F-v$ to compute $F$-dependent $P(v)$ using our theory. First, Fig.~\ref{figBlockYanagidaCross}A shows that with a single parameter ($d^+$) we can quantitatively fit the three $F-v$ curves. Using this value, $k_0^+$, $k_0^-$, and $\gamma_0$  we calculated the dependence of $P(v)$ on $F$ using the $F-v$ curve from \cite{Visscher1999} (Fig.~\ref{figBlockYanagidaCross}B) and the one from \cite{Carter2005} (Fig.~\ref{figBlockYanagidaCross}C). Although there are variations in the shapes of the $F-v$ curves the major conclusion of our study that $P(v)$ at $F \ne 0$ must exhibit bimodal behavior holds. 

\begin{table}[t]
\centering
\caption{Characteristic distances $d^+$ and $d^-$ for the fitting lines with different data sets in Fig.~\ref{figBlockYanagidaCross}A.}
\begin{tabular}{lcccccc}
 &        & $k^+_0$  & $k^-_0$  & $|d^+|$&$|d^+-d^-|$\\ [-0.25 em]
 \hline
 &\cite{Nishiyama2002}& $\SI{133.4 }{s^{-1}}$  &  $\SI{0.6}{s^{-1}}$   & 1.4 nm & 2.9 nm  \\
 &\cite{Visscher1999}& $\SI{133.4 }{s^{-1}}$  &  $\SI{0.6}{s^{-1}}$   & 0.6 nm & 3.0 nm  \\
 &\cite{Carter2005}&       $\SI{133.0 }{s^{-1}}$  &  $\SI{0.2}{s^{-1}}$   & 2.1 nm & 3.3 nm  \\
\end{tabular}
\label{tableratesT7}
\end{table} 

\begin{figure}
\centerline{\includegraphics[width=1\columnwidth]{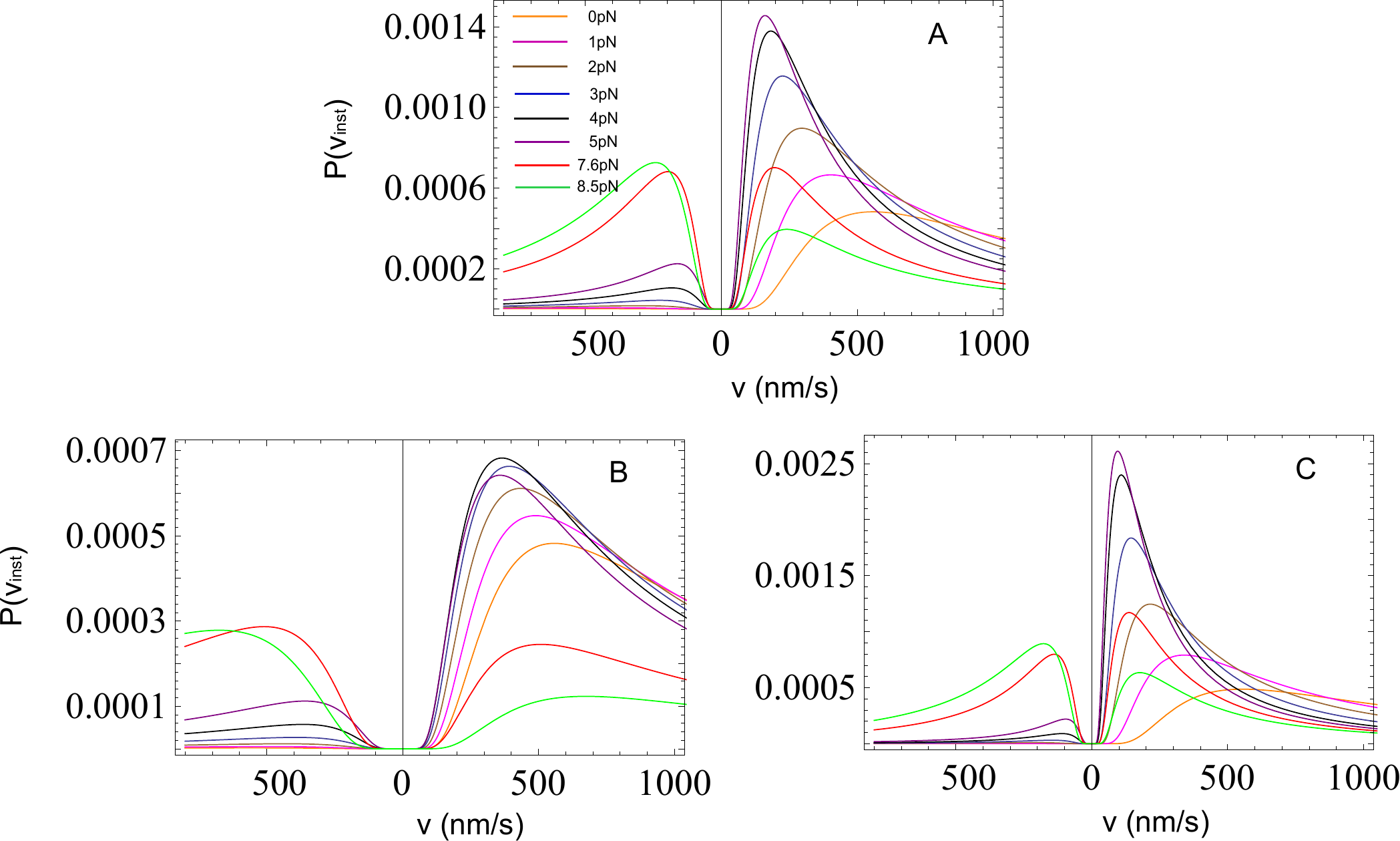}}
\caption{Predictions for $P(v_{inst})$ at different values of $F$ with the parameters listed in the first row (\textbf{A}), second row (\textbf{B}), and third row (\textbf{C}) in the Table~S1. }
\label{figpinstantv}
\end{figure}
{\bf Distribution of ``instantaneous" velocity $v_{inst}$ as a function of load:}
In several important studies \cite{Valleriani08EPL,Tsygankov07PRE,Linden07BJ} sequential models with multiple intermediate states (between 4 and 6) connecting  two successive target binding states have been developed to establish general characteristics of dwell time distribution functions, $P(\tau)$. From the numerical solution of an appropriate master equation corresponding to the sequential kinetic model, the $F$-dependent instantaneous mean velocity has been reported for Kin1 without explicitly considering detachment \cite{Valleriani08EPL}. Good agreement with experiments is found using the six state sequential model. The measured $P(\tau)$ decays exponentially at all $F$ (see Fig.~S2 in \cite{Carter2005}). This finding is a consequence of the absence of correlation between successive steps (dwell time $>>$ jump time) that the motors take on the underlying lattice. 

If $P(\tau)$  decays  exponentially, then for our model, which quantitatively describes all the experimental $F-v$ curves (Fig.~S7A), the distribution of $\tau$ can be written as $P(\tau)=k_T \emph{e}^{-k_T\tau}$. The total dwell time distribution alone cannot be used to separately calculate the distribution of dwell times when the motor takes predominantly forward or backward steps without an underlying model. The dwell time   distributions for forward and backward steps are, $P^+(\tau)=k^+\emph{e}^{-k_T\tau}$, and $P^-(\tau)=k^-\emph{e}^{-k_T\tau}$ where, in our simple model, $k_T=k^++k^-+\gamma$. All the rates are $F$-dependent. The distribution of instantaneous velocities $(v_{inst}>0)$ and $(v_{inst}<0)$ can be readily calculated using the kinetic scheme given in Fig.~1 in the main text. We obtain $P(v_{inst}>0)$ using,
$P(v_{inst}>0)=\int_0^\infty P^+(\tau) \delta(v_{inst} - \frac{1}{\tau}) d\tau$. Note that for convenience we measure $v$ in unit of step/s. We can convert the unit of velocity to nm/s by multiplying it with the step-size $s$, which is equal to $8.2$~nm for conventional kinesin, Kin1.   
Evaluating the integral leads to 
 \begin{eqnarray}
P(v_{inst}>0)&=&\frac{k^+}{v_{inst}^2} \emph{e}^{-k_T\frac{1}{v_{inst}}}.
\end{eqnarray}
A similar expression holds for $P(v_{inst}<0)$. 

The $F$-dependent $P(v_{inst})$ is shown in Fig.~S8 exhibits the bimodality  found in the distribution of the physical velocity discussed in the main text. However, $P(v_{inst})$ and $P(v)$ deviate qualitatively as $F$ decreases. In particular, $P(v_{inst})$ differs significantly from the expected Gaussian distribution at $F=0$ (Fig.~S8). From this analysis we draw two important conclusions: (1) From the measured $F$-dependent dwell time distributions alone, as reported in Fig.~S2 in \cite{Carter2005}, one cannot separate $P(v_{inst})$ into $P(v_{inst}>0)$ and $P(v_{inst}<0)$ without a tractable model that includes the possibility that the motor can detach from the track. The more complicated models \cite{Valleriani08EPL,Tsygankov07PRE,Linden07BJ} cannot produce expressions for $P(v_{inst}>0)$ and $P(v_{inst}<0)$, which can be used to analyze experimental data. In this context, the simple but accurate model that we have proposed, is ideally suited to interpret available data and also make testable predictions (see main text). (2) At zero $F$ where $k^-$ and $\gamma$ are negligible the velocity distribution (dominated by positive velocities) is well-approximated by a Gaussian (see Fig.~2 in the main text). That this is not the case for $P(v_{inst})$, but is found for $P(v)$ (see below), suggests that $v_{inst}$ may not be the correct way to model the motor velocity.  

{\bf Gaussian limit for  the velocity distribution:}
The velocity distribution $P(v)$ is typically fit using a Gaussian distribution based on the expectation that the motor takes $n$ steps with $k^-/k^+$ and $\gamma/k^+\ll 1$. If the steps are completely uncorrelated, as is the case in many motors then from central limit theorem it follows that $P(v)$ should approximately be a Gaussian provided $n\gg 1$. In order to address when this approximation is valid we consider a model with $k^-=\gamma=0$, which is appropriate at $F = 0$. Since the motor never detaches from the track in this limit, the velocity is $v=\frac{N}{t}$,
where the motor takes a fixed number of steps $n=N$ in each run.  The distribution of velocities for an ensemble of such chemically identical motors is  $P_N(v)$. %Note that for convenience we measure $v$ in unit of step/s. We can convert the unit of velocity to nm/s by multiplying it with the step-size $s$, which is equal to $8.2$~nm for conventional kinesin, Kin1.   
 
 We derive an exact expression for $P_N(v)$. The probability density $P(n,t_1,t_2,...,t_n)$ for the motor to take $n$ steps, with the first one at $t_1$, the second at $t_2$, and so on, is
$P(n,t_1,t_2,...,t_n)=[k^+ e^{-k^+ t_1}][k^+ e^{-k^+ (t_2-t_1)}]...[k^+ e^{-k^+ (t_n-t_{n-1})}]
=(k^+)^n e^{-k^+t_n}$
Thus, the joint probability density for the motor to jump $n$ steps in $t_n$ time is
\begin{eqnarray}
P(n,t_n)&=&\int_0^{t_n}P(n,t_1,t_2,...,t_n)dt_1\int_{t_1}^{t_n} dt_2... \int_{t_{n-2}}^{t_n} dt_{n-1}\\
&=&(k^+)^n e^{-k^+t_n}\frac{(t_n)^{n-1}}{(n-1)!},
\label{0Pnt}
\end{eqnarray}
from which the velocity distribution, may be obtained as, 
\begin{eqnarray}
P_N(v)&=&\int_{0}^{\infty}dt P(n=N,t)\delta\left(v-\frac{n}{t}\right)\\
&=&\frac{n}{v^2(n-1)!}\left(\frac{n}{v}\right)^{n-1} (k^+)^n \exp\left({-k^+\frac{n}{v}}\right) \Big\vert\frac{}{}_{n=N}
\label{0Pv}
\end{eqnarray}

\begin{figure}
\centerline{\includegraphics[width=1\columnwidth]{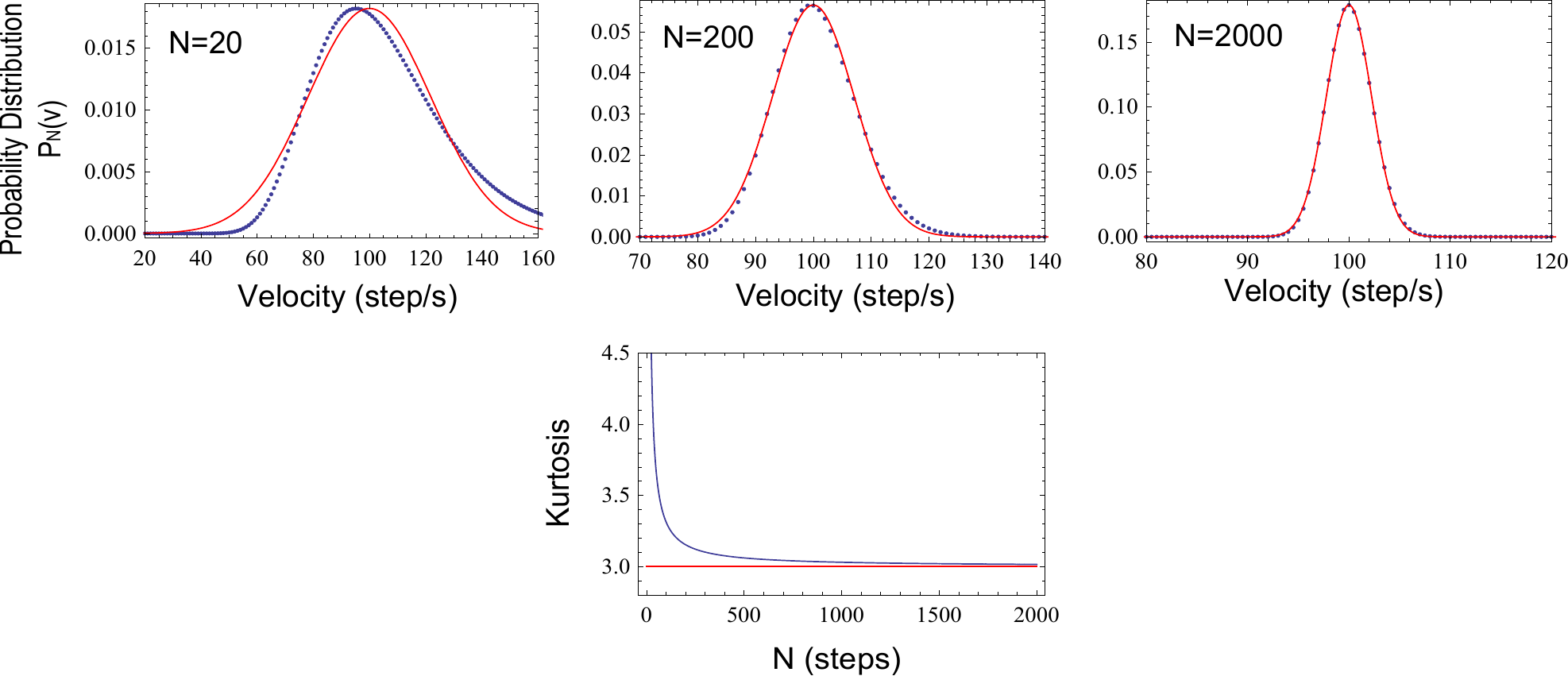}}
\caption{Comparision of theory and Kinetic Monte Carlo simulation in the model where the motor can only take forward step with a rate of $k^+=100$ /s. A: probability distribution of time for taking $N=200$ steps and B: probability distribution of velocity measured by all motors finishing $N=200$ steps. Red lines are predicted functions from the analytical model.}
\label{fignogama}
\end{figure}
The distributions $P_N(v)$s, plotted in Fig. \ref{fignogama}, for increasing values of $N$ using Eq.~\ref{0Pv}, show that as $N$ becomes larger, the Gaussian fit becomes increasingly accurate. For $N$=2000, the Gaussian is indistinguishable from the exact theory. We can also see that as $N$ increases, the kurtosis (defined as $\mu_4/\sigma^4$ where $\mu_4$ is the fourth moment about the mean and $\sigma$ is the standard deviation) value of $P_N(v)$ approaches 3, which is value for a Gaussian distribution function (see bottom panel in Fig. \ref{fignogama}. We surmise that for $N \approx100$ the velocity distribution should be well fit by a Gaussian when $\gamma=k^-=0$ in accord with Central Limit Theorem (see all Fig. 2 in the main text). 

\begin{figure}
\centerline{\includegraphics[width=\columnwidth]{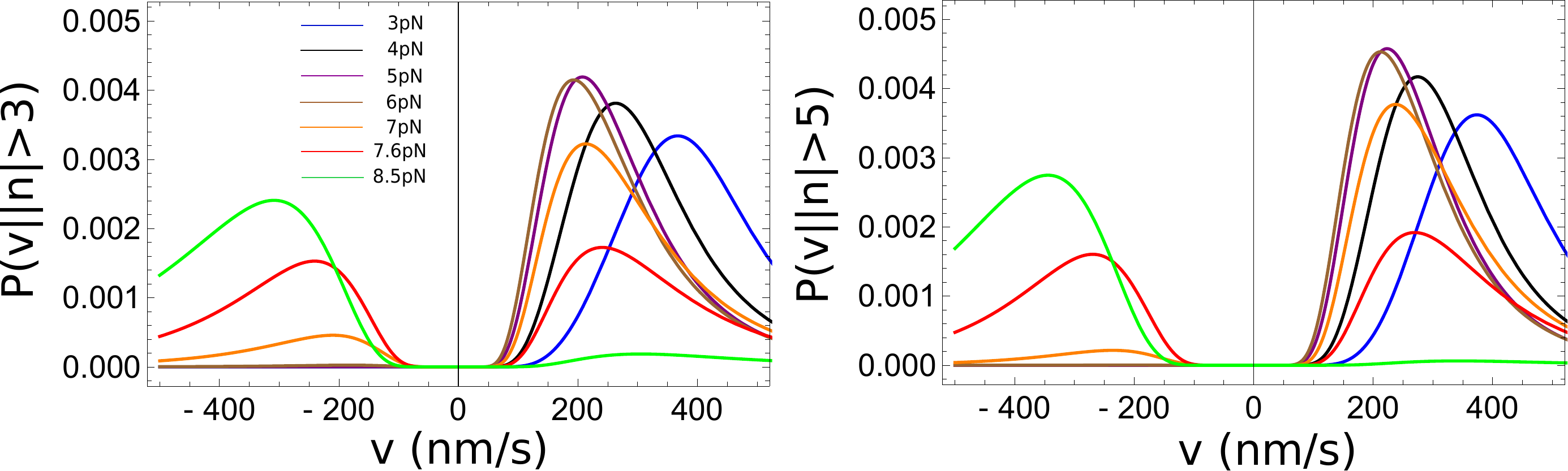}}
\caption{Force-dependent normalized velocity distribution obtained from motors that take more than $n_0$ steps. This restriction is imposed to assess if the bimodal distribution in $P(v)$ can be measured at $F \ne 0$.  The left (right) panel corresponds to $n_0 > 3$ ($n_0 > 5$). The preservation of bimodality shows that the main conclusions are robust.}
\end{figure}

 {\bf Randomness:}
Fluctuation effects in the reaction cycle of motors and more generally in any enzyme reaction can be succinctly captured using the randomness parameter ($r$) introduced by Schnitzer and Block \cite{Schnitzer1995}. For motors $r$ can be measured and computed  from the overall motor displacement $x(t)$ using \citep{Svoboda1994,Kolomeisky2000,Wang2003}
\begin{eqnarray}
r=lim_{t\to\infty}\frac{<x^2(t)>-<x(t)>^2}{s<x(t)>}=\frac{2D}{s<v>}.
\label{EqRandomness}
\end{eqnarray}
where $D$ is an effective diffusion constant. If there is no dispersion in the step-size $s$, $r$ can be used to obtain  the second moment ($r\equiv\frac{\sigma^2(\tau)}{<\tau>^2}$) of the dwell-time distribution $P(\tau)$ with $1/r$ setting the lower bound for the number of intermediate states needed to characterize  the reaction cycle \cite{Schnitzer1995}. If the step size is non-uniform then  $r=\frac{\sigma^2(s)}{<s>^2}+\frac{\sigma^2(\tau)}{<\tau>^2}$ where $\sigma^2(s)$ is the dispersion in $s$ \cite{Shaevitz2005}. Although the use of randomness parameter is not without problems \citep{Schnitzer1995,Wang2003} the ability to measure and analyze $r$ has given considerable insights into motor function.
 %However this approach has some limitations. Because $r$ only gives us information about second moments, not the full distributions, and many possible different distributions can have a same randomness, the interpretation of the results can prove difficult as the authors has been aware of in their papers \citep{Schnitzer1995,Wang2003}. 
 
\begin{figure}
\centerline{\includegraphics[width=0.6\columnwidth]{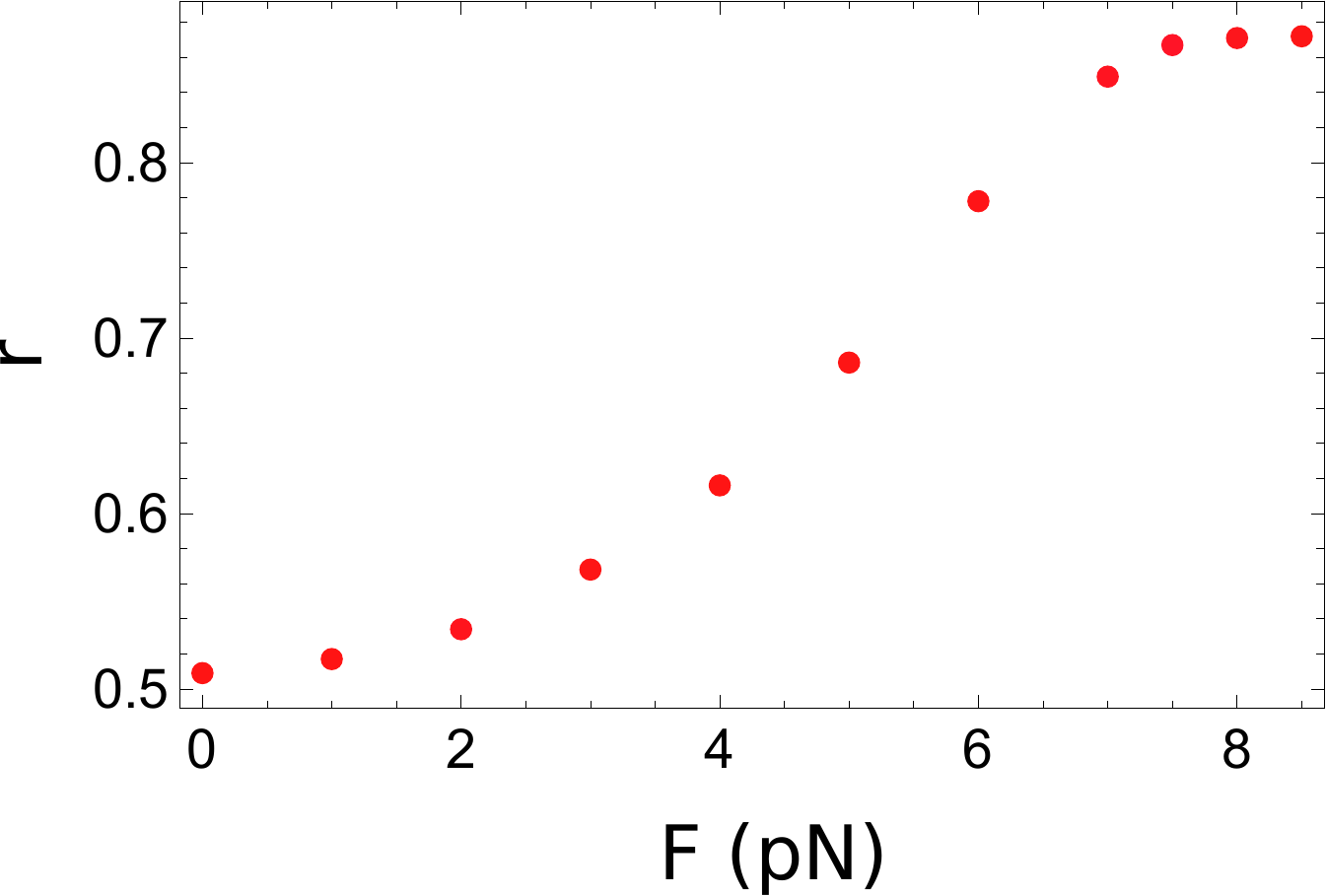}}
\caption{The dependence of the randomness parameter on $F$ for the model described in Fig. S6 A.}
\end{figure}

 In the simplified model considered in the main text the dwell-time distribution is $P(\tau)=k_T e^{-k_T\tau}$. Thus, $r$ is unity.  Even though $r=1$ the $P(v)$ distribution are highly $F$-dependent suggesting that $r$ alone is not indicator of the  non-Gaussian shape and bimodal structure of $P(v)$ at large forces. Nevertheless, $r = 1$ for the simple model does raise a question. Are  the central results in the main text is a consequence of the over simplification? Fig. S6B already shows that when an intermediate is considered $P(v)$ is bimodal with or without dispersion in the step size. We calculated the $F$-dependent $r$ parameter for the more realistic model in Fig. S6A. The results in Fig. 11 show that $r$ increases from about 0.5 at small forces to unity as $F$ approaches $F_s$. The values of $r$ are in qualitative agreement with experiments \cite{Visscher1999} and are satisfactory, given the difficulty in computing $r$ accurately even with models containing multiple intermediate states \cite{Fisher2001}. The combined results in Figs. S6 and S11 show that the central predictions in the main text hold even using models that qualitatively capture the $F$-dependent $r$ values.

{\bf Acknowledgements:} We are grateful to Steve Block, Changbong
Hyeon, and Jonathan Howard for useful discussions. This work was
supported in part by a grant from the National Science Foundation
(Grant No. CHE 13-61946).

%\bibliography{KMCKinesinRef}

\begin{thebibliography}{47}%
\makeatletter
\providecommand \@ifxundefined [1]{%
 \@ifx{#1\undefined}
}%
\providecommand \@ifnum [1]{%
 \ifnum #1\expandafter \@firstoftwo
 \else \expandafter \@secondoftwo
 \fi
}%
\providecommand \@ifx [1]{%
 \ifx #1\expandafter \@firstoftwo
 \else \expandafter \@secondoftwo
 \fi
}%
\providecommand \natexlab [1]{#1}%
\providecommand \enquote  [1]{``#1''}%
\providecommand \bibnamefont  [1]{#1}%
\providecommand \bibfnamefont [1]{#1}%
\providecommand \citenamefont [1]{#1}%
\providecommand \href@noop [0]{\@secondoftwo}%
\providecommand \href [0]{\begingroup \@sanitize@url \@href}%
\providecommand \@href[1]{\@@startlink{#1}\@@href}%
\providecommand \@@href[1]{\endgroup#1\@@endlink}%
\providecommand \@sanitize@url [0]{\catcode `\\12\catcode `\$12\catcode
  `\&12\catcode `\#12\catcode `\^12\catcode `\_12\catcode `\%12\relax}%
\providecommand \@@startlink[1]{}%
\providecommand \@@endlink[0]{}%
\providecommand \url  [0]{\begingroup\@sanitize@url \@url }%
\providecommand \@url [1]{\endgroup\@href {#1}{\urlprefix }}%
\providecommand \urlprefix  [0]{URL }%
\providecommand \Eprint [0]{\href }%
\providecommand \doibase [0]{http://dx.doi.org/}%
\providecommand \selectlanguage [0]{\@gobble}%
\providecommand \bibinfo  [0]{\@secondoftwo}%
\providecommand \bibfield  [0]{\@secondoftwo}%
\providecommand \translation [1]{[#1]}%
\providecommand \BibitemOpen [0]{}%
\providecommand \bibitemStop [0]{}%
\providecommand \bibitemNoStop [0]{.\EOS\space}%
\providecommand \EOS [0]{\spacefactor3000\relax}%
\providecommand \BibitemShut  [1]{\csname bibitem#1\endcsname}%
\let\auto@bib@innerbib\@empty
%</preamble>
\bibitem [{\citenamefont {Lodish}\ \emph {et~al.}(2007)\citenamefont {Lodish},
  \citenamefont {Berk}, \citenamefont {Kaiser}, \citenamefont {Krieger},
  \citenamefont {Scott}, \citenamefont {Bretscher}, \citenamefont {Ploegh},\
  and\ \citenamefont {Matsudaira}}]{bookLodish}%
  \BibitemOpen
  \bibfield  {author} {\bibinfo {author} {\bibfnamefont {H.~F.}\ \bibnamefont
  {Lodish}}, \bibinfo {author} {\bibfnamefont {A.}~\bibnamefont {Berk}},
  \bibinfo {author} {\bibfnamefont {C.~A.}\ \bibnamefont {Kaiser}}, \bibinfo
  {author} {\bibfnamefont {M.}~\bibnamefont {Krieger}}, \bibinfo {author}
  {\bibfnamefont {M.~P.}\ \bibnamefont {Scott}}, \bibinfo {author}
  {\bibfnamefont {A.}~\bibnamefont {Bretscher}}, \bibinfo {author}
  {\bibfnamefont {H.}~\bibnamefont {Ploegh}}, \ and\ \bibinfo {author}
  {\bibfnamefont {P.}~\bibnamefont {Matsudaira}},\ }\href@noop {} {\emph
  {\bibinfo {title} {Molecular Cell Biology}}},\ \bibinfo {edition} {sixth}\
  ed.\ (\bibinfo  {publisher} {Freeman},\ \bibinfo {address} {New York},\
  \bibinfo {year} {2007})\ p.\ \bibinfo {pages} {973}\BibitemShut {NoStop}%
\bibitem [{\citenamefont {Svoboda}\ and\ \citenamefont
  {Block}(1994)}]{Svoboda1994}%
  \BibitemOpen
  \bibfield  {author} {\bibinfo {author} {\bibfnamefont {K.}~\bibnamefont
  {Svoboda}}\ and\ \bibinfo {author} {\bibfnamefont {S.~M.}\ \bibnamefont
  {Block}},\ }\href@noop {} {\bibfield  {journal} {\bibinfo  {journal} {Cell}\
  }\textbf {\bibinfo {volume} {77}},\ \bibinfo {pages} {773} (\bibinfo {year}
  {1994})}\BibitemShut {NoStop}%
\bibitem [{\citenamefont {Schliwa}\ and\ \citenamefont
  {Woehlke}(2003)}]{schliwa2003}%
  \BibitemOpen
  \bibfield  {author} {\bibinfo {author} {\bibfnamefont {M.}~\bibnamefont
  {Schliwa}}\ and\ \bibinfo {author} {\bibfnamefont {G.}~\bibnamefont
  {Woehlke}},\ }\href@noop {} {\bibfield  {journal} {\bibinfo  {journal}
  {Nature}\ }\textbf {\bibinfo {volume} {422}},\ \bibinfo {pages} {759}
  (\bibinfo {year} {2003})}\BibitemShut {NoStop}%
\bibitem [{\citenamefont {Ray}\ \emph {et~al.}(1993)\citenamefont {Ray},
  \citenamefont {Meyhofer}, \citenamefont {Milligan},\ and\ \citenamefont
  {Howard}}]{Ray1993}%
  \BibitemOpen
  \bibfield  {author} {\bibinfo {author} {\bibfnamefont {S.}~\bibnamefont
  {Ray}}, \bibinfo {author} {\bibfnamefont {E.}~\bibnamefont {Meyhofer}},
  \bibinfo {author} {\bibfnamefont {R.~A.}\ \bibnamefont {Milligan}}, \ and\
  \bibinfo {author} {\bibfnamefont {J.}~\bibnamefont {Howard}},\ }\href@noop {}
  {\bibfield  {journal} {\bibinfo  {journal} {J. Cell Biol.}\ }\textbf
  {\bibinfo {volume} {121}},\ \bibinfo {pages} {1083} (\bibinfo {year}
  {1993})}\BibitemShut {NoStop}%
\bibitem [{\citenamefont {Spudich}\ and\ \citenamefont
  {Sivaramakrishnan}(2010)}]{Spudich2010}%
  \BibitemOpen
  \bibfield  {author} {\bibinfo {author} {\bibfnamefont {J.~A.}\ \bibnamefont
  {Spudich}}\ and\ \bibinfo {author} {\bibfnamefont {S.}~\bibnamefont
  {Sivaramakrishnan}},\ }\href@noop {} {\bibfield  {journal} {\bibinfo
  {journal} {Nat. Rev. Mol. Cell Bio.}\ }\textbf {\bibinfo {volume} {11}},\
  \bibinfo {pages} {128} (\bibinfo {year} {2010})}\BibitemShut {NoStop}%
\bibitem [{\citenamefont {Roberts}\ \emph {et~al.}(2013)\citenamefont
  {Roberts}, \citenamefont {Kon}, \citenamefont {Knight}, \citenamefont
  {Sutoh},\ and\ \citenamefont {Burgess}}]{Roberts2013}%
  \BibitemOpen
  \bibfield  {author} {\bibinfo {author} {\bibfnamefont {A.~J.}\ \bibnamefont
  {Roberts}}, \bibinfo {author} {\bibfnamefont {T.}~\bibnamefont {Kon}},
  \bibinfo {author} {\bibfnamefont {P.~J.}\ \bibnamefont {Knight}}, \bibinfo
  {author} {\bibfnamefont {K.}~\bibnamefont {Sutoh}}, \ and\ \bibinfo {author}
  {\bibfnamefont {S.~A.}\ \bibnamefont {Burgess}},\ }\href@noop {} {\bibfield
  {journal} {\bibinfo  {journal} {Nat. Rev. Mol. Cell Bio.}\ }\textbf {\bibinfo
  {volume} {14}},\ \bibinfo {pages} {713} (\bibinfo {year} {2013})}\BibitemShut
  {NoStop}%
\bibitem [{\citenamefont {Howard}(2001)}]{bookHoward}%
  \BibitemOpen
  \bibfield  {author} {\bibinfo {author} {\bibfnamefont {J.}~\bibnamefont
  {Howard}},\ }\href@noop {} {\emph {\bibinfo {title} {{Mechanics of Motor
  Proteins and the Cytoskeleton}}}}\ (\bibinfo  {publisher} {Sinauer
  Associates},\ \bibinfo {address} {Sunderland, Mass},\ \bibinfo {year}
  {2001})\BibitemShut {NoStop}%
\bibitem [{\citenamefont {Mallik}\ and\ \citenamefont
  {Gross}(2004)}]{Mallik2004}%
  \BibitemOpen
  \bibfield  {author} {\bibinfo {author} {\bibfnamefont {R.}~\bibnamefont
  {Mallik}}\ and\ \bibinfo {author} {\bibfnamefont {S.~P.}\ \bibnamefont
  {Gross}},\ }\href@noop {} {\bibfield  {journal} {\bibinfo  {journal} {Curr.
  Biol.}\ }\textbf {\bibinfo {volume} {14}},\ \bibinfo {pages} {R971} (\bibinfo
  {year} {2004})}\BibitemShut {NoStop}%
\bibitem [{\citenamefont {Lohman}\ \emph {et~al.}(2008)\citenamefont {Lohman},
  \citenamefont {Tomko},\ and\ \citenamefont {Wu}}]{Lohman2008}%
  \BibitemOpen
  \bibfield  {author} {\bibinfo {author} {\bibfnamefont {T.~M.}\ \bibnamefont
  {Lohman}}, \bibinfo {author} {\bibfnamefont {E.~J.}\ \bibnamefont {Tomko}}, \
  and\ \bibinfo {author} {\bibfnamefont {C.~G.}\ \bibnamefont {Wu}},\
  }\href@noop {} {\bibfield  {journal} {\bibinfo  {journal} {Nat. Rev. Mol.
  Cell Biol.}\ }\textbf {\bibinfo {volume} {9}},\ \bibinfo {pages} {391}
  (\bibinfo {year} {2008})}\BibitemShut {NoStop}%
\bibitem [{\citenamefont {Singleton}\ \emph {et~al.}(2007)\citenamefont
  {Singleton}, \citenamefont {Dillingham},\ and\ \citenamefont
  {Wigley}}]{Singleton2007}%
  \BibitemOpen
  \bibfield  {author} {\bibinfo {author} {\bibfnamefont {M.~R.}\ \bibnamefont
  {Singleton}}, \bibinfo {author} {\bibfnamefont {M.~S.}\ \bibnamefont
  {Dillingham}}, \ and\ \bibinfo {author} {\bibfnamefont {D.~B.}\ \bibnamefont
  {Wigley}},\ }\href@noop {} {\bibfield  {journal} {\bibinfo  {journal} {Annu.
  Rev. Biochem.}\ }\textbf {\bibinfo {volume} {76}},\ \bibinfo {pages} {23}
  (\bibinfo {year} {2007})}\BibitemShut {NoStop}%
\bibitem [{\citenamefont {Kojima}\ \emph {et~al.}(1997)\citenamefont {Kojima},
  \citenamefont {Muto}, \citenamefont {Higuchi},\ and\ \citenamefont
  {Yanagida}}]{Kojima1997}%
  \BibitemOpen
  \bibfield  {author} {\bibinfo {author} {\bibfnamefont {H.}~\bibnamefont
  {Kojima}}, \bibinfo {author} {\bibfnamefont {E.}~\bibnamefont {Muto}},
  \bibinfo {author} {\bibfnamefont {H.}~\bibnamefont {Higuchi}}, \ and\
  \bibinfo {author} {\bibfnamefont {T.}~\bibnamefont {Yanagida}},\ }\href@noop
  {} {\bibfield  {journal} {\bibinfo  {journal} {Biophys. J.}\ }\textbf
  {\bibinfo {volume} {73}},\ \bibinfo {pages} {2012} (\bibinfo {year}
  {1997})}\BibitemShut {NoStop}%
\bibitem [{\citenamefont {Nishiyama}\ \emph {et~al.}(2002)\citenamefont
  {Nishiyama}, \citenamefont {Higuchi},\ and\ \citenamefont
  {Yanagida}}]{Nishiyama2002}%
  \BibitemOpen
  \bibfield  {author} {\bibinfo {author} {\bibfnamefont {M.}~\bibnamefont
  {Nishiyama}}, \bibinfo {author} {\bibfnamefont {H.}~\bibnamefont {Higuchi}},
  \ and\ \bibinfo {author} {\bibfnamefont {T.}~\bibnamefont {Yanagida}},\
  }\href@noop {} {\bibfield  {journal} {\bibinfo  {journal} {Nat. Cell Biol.}\
  }\textbf {\bibinfo {volume} {4}},\ \bibinfo {pages} {790} (\bibinfo {year}
  {2002})}\BibitemShut {NoStop}%
\bibitem [{\citenamefont {Walter}\ \emph {et~al.}(2012)\citenamefont {Walter},
  \citenamefont {Beranek}, \citenamefont {Fischermeier},\ and\ \citenamefont
  {Diez}}]{Fischermeier2012}%
  \BibitemOpen
  \bibfield  {author} {\bibinfo {author} {\bibfnamefont {W.~J.}\ \bibnamefont
  {Walter}}, \bibinfo {author} {\bibfnamefont {V.}~\bibnamefont {Beranek}},
  \bibinfo {author} {\bibfnamefont {E.}~\bibnamefont {Fischermeier}}, \ and\
  \bibinfo {author} {\bibfnamefont {S.}~\bibnamefont {Diez}},\ }\href@noop {}
  {\bibfield  {journal} {\bibinfo  {journal} {PLoS One}\ }\textbf {\bibinfo
  {volume} {7}},\ \bibinfo {pages} {e42218} (\bibinfo {year}
  {2012})}\BibitemShut {NoStop}%
\bibitem [{\citenamefont {Courty}\ \emph {et~al.}(2006)\citenamefont {Courty},
  \citenamefont {Luccardini}, \citenamefont {Bellaiche}, \citenamefont
  {Cappello},\ and\ \citenamefont {Dahan}}]{Cappello2006}%
  \BibitemOpen
  \bibfield  {author} {\bibinfo {author} {\bibfnamefont {S.}~\bibnamefont
  {Courty}}, \bibinfo {author} {\bibfnamefont {C.}~\bibnamefont {Luccardini}},
  \bibinfo {author} {\bibfnamefont {Y.}~\bibnamefont {Bellaiche}}, \bibinfo
  {author} {\bibfnamefont {G.}~\bibnamefont {Cappello}}, \ and\ \bibinfo
  {author} {\bibfnamefont {M.}~\bibnamefont {Dahan}},\ }\href@noop {}
  {\bibfield  {journal} {\bibinfo  {journal} {Nano Lett.}\ }\textbf {\bibinfo
  {volume} {6}},\ \bibinfo {pages} {1491} (\bibinfo {year} {2006})}\BibitemShut
  {NoStop}%
\bibitem [{\citenamefont {Uemura}\ \emph {et~al.}(2002)\citenamefont {Uemura},
  \citenamefont {Kawaguchi}, \citenamefont {Yajima}, \citenamefont {Edamatsu},
  \citenamefont {Toyoshima},\ and\ \citenamefont {Ishiwata}}]{Uemura2002}%
  \BibitemOpen
  \bibfield  {author} {\bibinfo {author} {\bibfnamefont {S.}~\bibnamefont
  {Uemura}}, \bibinfo {author} {\bibfnamefont {K.}~\bibnamefont {Kawaguchi}},
  \bibinfo {author} {\bibfnamefont {J.}~\bibnamefont {Yajima}}, \bibinfo
  {author} {\bibfnamefont {M.}~\bibnamefont {Edamatsu}}, \bibinfo {author}
  {\bibfnamefont {Y.~Y.}\ \bibnamefont {Toyoshima}}, \ and\ \bibinfo {author}
  {\bibfnamefont {S.}~\bibnamefont {Ishiwata}},\ }\href@noop {} {\bibfield
  {journal} {\bibinfo  {journal} {Proc. Natl. Acad. Sci. U.S.A.}\ }\textbf
  {\bibinfo {volume} {99}},\ \bibinfo {pages} {3} (\bibinfo {year}
  {2002})}\BibitemShut {NoStop}%
\bibitem [{\citenamefont {Svoboda}\ \emph {et~al.}(1993)\citenamefont
  {Svoboda}, \citenamefont {Schmidt}, \citenamefont {Schnapp},\ and\
  \citenamefont {Block}}]{Svoboda1993}%
  \BibitemOpen
  \bibfield  {author} {\bibinfo {author} {\bibfnamefont {K.}~\bibnamefont
  {Svoboda}}, \bibinfo {author} {\bibfnamefont {C.~F.}\ \bibnamefont
  {Schmidt}}, \bibinfo {author} {\bibfnamefont {B.~J.}\ \bibnamefont
  {Schnapp}}, \ and\ \bibinfo {author} {\bibfnamefont {S.~M.}\ \bibnamefont
  {Block}},\ }\href@noop {} {\bibfield  {journal} {\bibinfo  {journal}
  {Nature}\ }\textbf {\bibinfo {volume} {365}},\ \bibinfo {pages} {721}
  (\bibinfo {year} {1993})}\BibitemShut {NoStop}%
\bibitem [{\citenamefont {Fehr}\ \emph {et~al.}(2008)\citenamefont {Fehr},
  \citenamefont {Asbury},\ and\ \citenamefont {Block}}]{Fehr2008}%
  \BibitemOpen
  \bibfield  {author} {\bibinfo {author} {\bibfnamefont {A.~N.}\ \bibnamefont
  {Fehr}}, \bibinfo {author} {\bibfnamefont {C.~L.}\ \bibnamefont {Asbury}}, \
  and\ \bibinfo {author} {\bibfnamefont {S.~M.}\ \bibnamefont {Block}},\
  }\href@noop {} {\bibfield  {journal} {\bibinfo  {journal} {Biophys. J.}\
  }\textbf {\bibinfo {volume} {94}},\ \bibinfo {pages} {L20} (\bibinfo {year}
  {2008})}\BibitemShut {NoStop}%
\bibitem [{\citenamefont {Schnitzer}\ \emph {et~al.}(2000)\citenamefont
  {Schnitzer}, \citenamefont {Visscher},\ and\ \citenamefont
  {Block}}]{Schnitzer2000}%
  \BibitemOpen
  \bibfield  {author} {\bibinfo {author} {\bibfnamefont {M.~J.}\ \bibnamefont
  {Schnitzer}}, \bibinfo {author} {\bibfnamefont {K.}~\bibnamefont {Visscher}},
  \ and\ \bibinfo {author} {\bibfnamefont {S.~M.}\ \bibnamefont {Block}},\
  }\href@noop {} {\bibfield  {journal} {\bibinfo  {journal} {Nat. Cell Biol.}\
  }\textbf {\bibinfo {volume} {2}},\ \bibinfo {pages} {718} (\bibinfo {year}
  {2000})}\BibitemShut {NoStop}%
\bibitem [{\citenamefont {Milic}\ \emph {et~al.}(2014)\citenamefont {Milic},
  \citenamefont {Andreasson}, \citenamefont {Hancock},\ and\ \citenamefont
  {Block}}]{Milic14PNAS}%
  \BibitemOpen
  \bibfield  {author} {\bibinfo {author} {\bibfnamefont {B.}~\bibnamefont
  {Milic}}, \bibinfo {author} {\bibfnamefont {J.~O.~L.}\ \bibnamefont
  {Andreasson}}, \bibinfo {author} {\bibfnamefont {W.~O.}\ \bibnamefont
  {Hancock}}, \ and\ \bibinfo {author} {\bibfnamefont {S.~M.}\ \bibnamefont
  {Block}},\ }\href@noop {} {\bibfield  {journal} {\bibinfo  {journal} {{Proc.
  Natl. Acad. Sci.}}\ }\textbf {\bibinfo {volume} {111}},\ \bibinfo {pages}
  {14136} (\bibinfo {year} {2014})}\BibitemShut {NoStop}%
\bibitem [{\citenamefont {Kolomeisky}\ and\ \citenamefont
  {Fisher}(2007)}]{Kolomeisky2007}%
  \BibitemOpen
  \bibfield  {author} {\bibinfo {author} {\bibfnamefont {A.~B.}\ \bibnamefont
  {Kolomeisky}}\ and\ \bibinfo {author} {\bibfnamefont {M.~E.}\ \bibnamefont
  {Fisher}},\ }\href@noop {} {\bibfield  {journal} {\bibinfo  {journal} {Annu.
  Rev. Phys. Chem.}\ }\textbf {\bibinfo {volume} {58}},\ \bibinfo {pages} {675}
  (\bibinfo {year} {2007})}\BibitemShut {NoStop}%
\bibitem [{\citenamefont {Schnitzer}\ and\ \citenamefont
  {Block}(1995)}]{Schnitzer1995}%
  \BibitemOpen
  \bibfield  {author} {\bibinfo {author} {\bibfnamefont {M.~J.}\ \bibnamefont
  {Schnitzer}}\ and\ \bibinfo {author} {\bibfnamefont {S.~M.}\ \bibnamefont
  {Block}},\ }\href@noop {} {\bibfield  {journal} {\bibinfo  {journal} {Cold
  Spring Harb. Sym.}\ }\textbf {\bibinfo {volume} {60}},\ \bibinfo {pages} {793
  } (\bibinfo {year} {1995})}\BibitemShut {NoStop}%
\bibitem [{\citenamefont {Taniguchi}\ \emph {et~al.}(2005)\citenamefont
  {Taniguchi}, \citenamefont {Nishiyama}, \citenamefont {Ishii},\ and\
  \citenamefont {Yanagida}}]{Taniguchi2005}%
  \BibitemOpen
  \bibfield  {author} {\bibinfo {author} {\bibfnamefont {Y.}~\bibnamefont
  {Taniguchi}}, \bibinfo {author} {\bibfnamefont {M.}~\bibnamefont
  {Nishiyama}}, \bibinfo {author} {\bibfnamefont {Y.}~\bibnamefont {Ishii}}, \
  and\ \bibinfo {author} {\bibfnamefont {T.}~\bibnamefont {Yanagida}},\
  }\href@noop {} {\bibfield  {journal} {\bibinfo  {journal} {Nat. Chem. Biol.}\
  }\textbf {\bibinfo {volume} {1}},\ \bibinfo {pages} {342} (\bibinfo {year}
  {2005})}\BibitemShut {NoStop}%
\bibitem [{\citenamefont {Shaevitz}\ \emph {et~al.}(2005)\citenamefont
  {Shaevitz}, \citenamefont {Block},\ and\ \citenamefont
  {Schnitzer}}]{Shaevitz2005}%
  \BibitemOpen
  \bibfield  {author} {\bibinfo {author} {\bibfnamefont {J.~W.}\ \bibnamefont
  {Shaevitz}}, \bibinfo {author} {\bibfnamefont {S.~M.}\ \bibnamefont {Block}},
  \ and\ \bibinfo {author} {\bibfnamefont {M.~J.}\ \bibnamefont {Schnitzer}},\
  }\href@noop {} {\bibfield  {journal} {\bibinfo  {journal} {Biophys. J.}\
  }\textbf {\bibinfo {volume} {89}},\ \bibinfo {pages} {2277} (\bibinfo {year}
  {2005})}\BibitemShut {NoStop}%
\bibitem [{\citenamefont {Chowdhury}(2013)}]{Chowdhury2013}%
  \BibitemOpen
  \bibfield  {author} {\bibinfo {author} {\bibfnamefont {D.}~\bibnamefont
  {Chowdhury}},\ }\href@noop {} {\bibfield  {journal} {\bibinfo  {journal}
  {Phys. Rep.}\ }\textbf {\bibinfo {volume} {529}},\ \bibinfo {pages} {1}
  (\bibinfo {year} {2013})}\BibitemShut {NoStop}%
\bibitem [{\citenamefont {Fisher}\ and\ \citenamefont
  {Kolomeisky}(2001)}]{Fisher2001}%
  \BibitemOpen
  \bibfield  {author} {\bibinfo {author} {\bibfnamefont {M.~E.}\ \bibnamefont
  {Fisher}}\ and\ \bibinfo {author} {\bibfnamefont {A.~B.}\ \bibnamefont
  {Kolomeisky}},\ }\href@noop {} {\bibfield  {journal} {\bibinfo  {journal}
  {Proc. Natl. Acad. Sci. U.S.A.}\ }\textbf {\bibinfo {volume} {98}},\ \bibinfo
  {pages} {7748} (\bibinfo {year} {2001})}\BibitemShut {NoStop}%
\bibitem [{\citenamefont {Ali}\ \emph {et~al.}(2008)\citenamefont {Ali},
  \citenamefont {Lu}, \citenamefont {Bookwalter}, \citenamefont {Warshaw},\
  and\ \citenamefont {Trybus}}]{Ali2008}%
  \BibitemOpen
  \bibfield  {author} {\bibinfo {author} {\bibfnamefont {M.~Y.}\ \bibnamefont
  {Ali}}, \bibinfo {author} {\bibfnamefont {H.}~\bibnamefont {Lu}}, \bibinfo
  {author} {\bibfnamefont {C.~S.}\ \bibnamefont {Bookwalter}}, \bibinfo
  {author} {\bibfnamefont {D.~M.}\ \bibnamefont {Warshaw}}, \ and\ \bibinfo
  {author} {\bibfnamefont {K.~M.}\ \bibnamefont {Trybus}},\ }\href@noop {}
  {\bibfield  {journal} {\bibinfo  {journal} {Proc. Natl. Acad. Sci. U.S.A.}\
  }\textbf {\bibinfo {volume} {105}},\ \bibinfo {pages} {4691} (\bibinfo {year}
  {2008})}\BibitemShut {NoStop}%
\bibitem [{\citenamefont {Soppina}\ \emph {et~al.}(2014)\citenamefont
  {Soppina}, \citenamefont {Norris}, \citenamefont {Dizaji}, \citenamefont
  {Kortus}, \citenamefont {Veatch}, \citenamefont {Peckham},\ and\
  \citenamefont {Verhey}}]{Soppina2014}%
  \BibitemOpen
  \bibfield  {author} {\bibinfo {author} {\bibfnamefont {V.}~\bibnamefont
  {Soppina}}, \bibinfo {author} {\bibfnamefont {S.~R.}\ \bibnamefont {Norris}},
  \bibinfo {author} {\bibfnamefont {A.~S.}\ \bibnamefont {Dizaji}}, \bibinfo
  {author} {\bibfnamefont {M.}~\bibnamefont {Kortus}}, \bibinfo {author}
  {\bibfnamefont {S.}~\bibnamefont {Veatch}}, \bibinfo {author} {\bibfnamefont
  {M.}~\bibnamefont {Peckham}}, \ and\ \bibinfo {author} {\bibfnamefont
  {K.~J.}\ \bibnamefont {Verhey}},\ }\href@noop {} {\bibfield  {journal}
  {\bibinfo  {journal} {Proc. Natl. Acad. Sci. U.S.A.}\ }\textbf {\bibinfo
  {volume} {111}},\ \bibinfo {pages} {5562} (\bibinfo {year}
  {2014})}\BibitemShut {NoStop}%
\bibitem [{\citenamefont {Xu}\ \emph {et~al.}(2013)\citenamefont {Xu},
  \citenamefont {King}, \citenamefont {Lapierre-landry},\ and\ \citenamefont
  {Nemec}}]{Xu2013}%
  \BibitemOpen
  \bibfield  {author} {\bibinfo {author} {\bibfnamefont {J.}~\bibnamefont
  {Xu}}, \bibinfo {author} {\bibfnamefont {S.~J.}\ \bibnamefont {King}},
  \bibinfo {author} {\bibfnamefont {M.}~\bibnamefont {Lapierre-landry}}, \ and\
  \bibinfo {author} {\bibfnamefont {B.}~\bibnamefont {Nemec}},\ }\href@noop {}
  {\bibfield  {journal} {\bibinfo  {journal} {Biophys. J.}\ }\textbf {\bibinfo
  {volume} {105}},\ \bibinfo {pages} {L23} (\bibinfo {year}
  {2013})}\BibitemShut {NoStop}%
\bibitem [{\citenamefont {Hammond}\ \emph {et~al.}(2009)\citenamefont
  {Hammond}, \citenamefont {Cai}, \citenamefont {Blasius}, \citenamefont {Li},
  \citenamefont {Jiang}, \citenamefont {Jih}, \citenamefont {Meyhofer},\ and\
  \citenamefont {Verhey}}]{Hammond2009}%
  \BibitemOpen
  \bibfield  {author} {\bibinfo {author} {\bibfnamefont {J.~W.}\ \bibnamefont
  {Hammond}}, \bibinfo {author} {\bibfnamefont {D.}~\bibnamefont {Cai}},
  \bibinfo {author} {\bibfnamefont {T.~L.}\ \bibnamefont {Blasius}}, \bibinfo
  {author} {\bibfnamefont {Z.}~\bibnamefont {Li}}, \bibinfo {author}
  {\bibfnamefont {Y.}~\bibnamefont {Jiang}}, \bibinfo {author} {\bibfnamefont
  {G.~T.}\ \bibnamefont {Jih}}, \bibinfo {author} {\bibfnamefont
  {E.}~\bibnamefont {Meyhofer}}, \ and\ \bibinfo {author} {\bibfnamefont
  {K.~J.}\ \bibnamefont {Verhey}},\ }\href@noop {} {\bibfield  {journal}
  {\bibinfo  {journal} {PLoS Biol.}\ }\textbf {\bibinfo {volume} {7}} (\bibinfo
  {year} {2009})}\BibitemShut {NoStop}%
\bibitem [{\citenamefont {M\"{u}ller}\ \emph {et~al.}(2010)\citenamefont
  {M\"{u}ller}, \citenamefont {Berger}, \citenamefont {Klumpp},\ and\
  \citenamefont {Lipowsky}}]{bookdetachmentrate}%
  \BibitemOpen
  \bibfield  {author} {\bibinfo {author} {\bibfnamefont {M.~J.~I.}\
  \bibnamefont {M\"{u}ller}}, \bibinfo {author} {\bibfnamefont
  {F.}~\bibnamefont {Berger}}, \bibinfo {author} {\bibfnamefont
  {S.}~\bibnamefont {Klumpp}}, \ and\ \bibinfo {author} {\bibfnamefont
  {R.}~\bibnamefont {Lipowsky}},\ }\enquote {\bibinfo {title} {Cargo transport
  by teams of molecular motors: Basic mechanisms for intracellular drug
  delivery},}\ in\ \href@noop {} {\emph {\bibinfo {booktitle}
  {Organelle-Specific Pharmaceutical Nanotechnology}}}\ (\bibinfo  {publisher}
  {John Wiley \& Sons, Inc.},\ \bibinfo {year} {2010})\ pp.\ \bibinfo {pages}
  {289--309}\BibitemShut {NoStop}%
\bibitem [{\citenamefont {Valleriani}\ \emph {et~al.}(2008)\citenamefont
  {Valleriani}, \citenamefont {Liepelt},\ and\ \citenamefont
  {Lipowsky}}]{Valleriani08EPL}%
  \BibitemOpen
  \bibfield  {author} {\bibinfo {author} {\bibfnamefont {A.}~\bibnamefont
  {Valleriani}}, \bibinfo {author} {\bibfnamefont {S.}~\bibnamefont {Liepelt}},
  \ and\ \bibinfo {author} {\bibfnamefont {R.}~\bibnamefont {Lipowsky}},\
  }\href@noop {} {\bibfield  {journal} {\bibinfo  {journal} {{Euro. Phys.
  Lett.}}\ }\textbf {\bibinfo {volume} {82}},\ \bibinfo {pages} {28011}
  (\bibinfo {year} {2008})}\BibitemShut {NoStop}%
\bibitem [{\citenamefont {Tsygankov}\ \emph {et~al.}(2007)\citenamefont
  {Tsygankov}, \citenamefont {Linden},\ and\ \citenamefont
  {Fisher}}]{Tsygankov07PRE}%
  \BibitemOpen
  \bibfield  {author} {\bibinfo {author} {\bibfnamefont {D.}~\bibnamefont
  {Tsygankov}}, \bibinfo {author} {\bibfnamefont {M.}~\bibnamefont {Linden}}, \
  and\ \bibinfo {author} {\bibfnamefont {M.~E.}\ \bibnamefont {Fisher}},\
  }\href@noop {} {\bibfield  {journal} {\bibinfo  {journal} {{Phys. Rev. E}}\
  }\textbf {\bibinfo {volume} {75}},\ \bibinfo {pages} {021909} (\bibinfo
  {year} {2007})}\BibitemShut {NoStop}%
\bibitem [{\citenamefont {Linden}\ and\ \citenamefont
  {Wallin}(2007)}]{Linden07BJ}%
  \BibitemOpen
  \bibfield  {author} {\bibinfo {author} {\bibfnamefont {M.}~\bibnamefont
  {Linden}}\ and\ \bibinfo {author} {\bibfnamefont {M.}~\bibnamefont
  {Wallin}},\ }\href@noop {} {\bibfield  {journal} {\bibinfo  {journal}
  {Biophys. J.}\ }\textbf {\bibinfo {volume} {92}},\ \bibinfo {pages} {3804}
  (\bibinfo {year} {2007})}\BibitemShut {NoStop}%
\bibitem [{\citenamefont {Carter}\ and\ \citenamefont
  {Cross}(2005)}]{Carter2005}%
  \BibitemOpen
  \bibfield  {author} {\bibinfo {author} {\bibfnamefont {N.~J.}\ \bibnamefont
  {Carter}}\ and\ \bibinfo {author} {\bibfnamefont {R.~A.}\ \bibnamefont
  {Cross}},\ }\href@noop {} {\bibfield  {journal} {\bibinfo  {journal}
  {Nature}\ }\textbf {\bibinfo {volume} {435}},\ \bibinfo {pages} {308}
  (\bibinfo {year} {2005})}\BibitemShut {NoStop}%
\bibitem [{\citenamefont {Efron}\ and\ \citenamefont
  {Tibshirani}(1993)}]{bookbootstrap}%
  \BibitemOpen
  \bibfield  {author} {\bibinfo {author} {\bibfnamefont {B.}~\bibnamefont
  {Efron}}\ and\ \bibinfo {author} {\bibfnamefont {R.~J.}\ \bibnamefont
  {Tibshirani}},\ }\enquote {\bibinfo {title} {An introduction to the
  bootstrap},}\ \ (\bibinfo  {publisher} {Chapman \& Hall},\ \bibinfo {address}
  {London},\ \bibinfo {year} {1993})\BibitemShut {NoStop}%
\bibitem [{\citenamefont {Vale}\ \emph {et~al.}(1985)\citenamefont {Vale},
  \citenamefont {Reese},\ and\ \citenamefont {Sheetz}}]{Vale1985}%
  \BibitemOpen
  \bibfield  {author} {\bibinfo {author} {\bibfnamefont {R.~D.}\ \bibnamefont
  {Vale}}, \bibinfo {author} {\bibfnamefont {T.~S.}\ \bibnamefont {Reese}}, \
  and\ \bibinfo {author} {\bibfnamefont {M.~P.}\ \bibnamefont {Sheetz}},\
  }\href@noop {} {\bibfield  {journal} {\bibinfo  {journal} {Cell}\ }\textbf
  {\bibinfo {volume} {42}},\ \bibinfo {pages} {39} (\bibinfo {year}
  {1985})}\BibitemShut {NoStop}%
\bibitem [{\citenamefont {Yildiz}\ \emph {et~al.}(2004)\citenamefont {Yildiz},
  \citenamefont {Tomishige}, \citenamefont {Vale},\ and\ \citenamefont
  {Selvin}}]{Yildiz2004a}%
  \BibitemOpen
  \bibfield  {author} {\bibinfo {author} {\bibfnamefont {A.}~\bibnamefont
  {Yildiz}}, \bibinfo {author} {\bibfnamefont {M.}~\bibnamefont {Tomishige}},
  \bibinfo {author} {\bibfnamefont {R.~D.}\ \bibnamefont {Vale}}, \ and\
  \bibinfo {author} {\bibfnamefont {P.~R.}\ \bibnamefont {Selvin}},\
  }\href@noop {} {\bibfield  {journal} {\bibinfo  {journal} {Sciences}\
  }\textbf {\bibinfo {volume} {303}},\ \bibinfo {pages} {676} (\bibinfo {year}
  {2004})}\BibitemShut {NoStop}%
\bibitem [{\citenamefont {Asbury}\ \emph {et~al.}(2003)\citenamefont {Asbury},
  \citenamefont {Fehr},\ and\ \citenamefont {Block}}]{Asbury2003}%
  \BibitemOpen
  \bibfield  {author} {\bibinfo {author} {\bibfnamefont {C.~L.}\ \bibnamefont
  {Asbury}}, \bibinfo {author} {\bibfnamefont {A.~N.}\ \bibnamefont {Fehr}}, \
  and\ \bibinfo {author} {\bibfnamefont {S.~M.}\ \bibnamefont {Block}},\
  }\href@noop {} {\bibfield  {journal} {\bibinfo  {journal} {Science}\ }\textbf
  {\bibinfo {volume} {302}},\ \bibinfo {pages} {2130} (\bibinfo {year}
  {2003})}\BibitemShut {NoStop}%
\bibitem [{\citenamefont {Coy}\ \emph {et~al.}(1999)\citenamefont {Coy},
  \citenamefont {Wagenbach},\ and\ \citenamefont {Howard}}]{Coy1999}%
  \BibitemOpen
  \bibfield  {author} {\bibinfo {author} {\bibfnamefont {D.~L.}\ \bibnamefont
  {Coy}}, \bibinfo {author} {\bibfnamefont {M.}~\bibnamefont {Wagenbach}}, \
  and\ \bibinfo {author} {\bibfnamefont {J.}~\bibnamefont {Howard}},\
  }\href@noop {} {\bibfield  {journal} {\bibinfo  {journal} {J. Biol. Chem.}\
  }\textbf {\bibinfo {volume} {274}},\ \bibinfo {pages} {3667} (\bibinfo {year}
  {1999})}\BibitemShut {NoStop}%
\bibitem [{\citenamefont {Hyeon}\ \emph {et~al.}(2009)\citenamefont {Hyeon},
  \citenamefont {Klumpp},\ and\ \citenamefont {Onuchic}}]{Hyeon09PCCP}%
  \BibitemOpen
  \bibfield  {author} {\bibinfo {author} {\bibfnamefont {C.}~\bibnamefont
  {Hyeon}}, \bibinfo {author} {\bibfnamefont {S.}~\bibnamefont {Klumpp}}, \
  and\ \bibinfo {author} {\bibfnamefont {J.~N.}\ \bibnamefont {Onuchic}},\
  }\href@noop {} {\bibfield  {journal} {\bibinfo  {journal} {{Phys. Chem. Chem.
  Phys.}}\ }\textbf {\bibinfo {volume} {11}},\ \bibinfo {pages} {4899}
  (\bibinfo {year} {2009})}\BibitemShut {NoStop}%
\bibitem [{\citenamefont {Hughes}\ \emph {et~al.}(2013)\citenamefont {Hughes},
  \citenamefont {Shastry}, \citenamefont {Hancock},\ and\ \citenamefont
  {Fricks}}]{Hughes2013}%
  \BibitemOpen
  \bibfield  {author} {\bibinfo {author} {\bibfnamefont {J.}~\bibnamefont
  {Hughes}}, \bibinfo {author} {\bibfnamefont {S.}~\bibnamefont {Shastry}},
  \bibinfo {author} {\bibfnamefont {W.~O.}\ \bibnamefont {Hancock}}, \ and\
  \bibinfo {author} {\bibfnamefont {J.}~\bibnamefont {Fricks}},\ }\href@noop {}
  {\bibfield  {journal} {\bibinfo  {journal} {J. Agr. Biol. Envir. S.}\
  }\textbf {\bibinfo {volume} {18}},\ \bibinfo {pages} {204} (\bibinfo {year}
  {2013})}\BibitemShut {NoStop}%
\bibitem [{\citenamefont {Bortz}\ \emph {et~al.}(1975)\citenamefont {Bortz},
  \citenamefont {Kalos},\ and\ \citenamefont {Lebowitz}}]{Bortz75JCP}%
  \BibitemOpen
  \bibfield  {author} {\bibinfo {author} {\bibfnamefont {A.}~\bibnamefont
  {Bortz}}, \bibinfo {author} {\bibfnamefont {M.}~\bibnamefont {Kalos}}, \ and\
  \bibinfo {author} {\bibfnamefont {J.}~\bibnamefont {Lebowitz}},\ }\href@noop
  {} {\bibfield  {journal} {\bibinfo  {journal} {{J. Chem. Phys.}}\ }\textbf
  {\bibinfo {volume} {17}},\ \bibinfo {pages} {10} (\bibinfo {year}
  {1975})}\BibitemShut {NoStop}%
\bibitem [{\citenamefont {Visscher}\ \emph {et~al.}(1999)\citenamefont
  {Visscher}, \citenamefont {Schnitzer},\ and\ \citenamefont
  {Block}}]{Visscher1999}%
  \BibitemOpen
  \bibfield  {author} {\bibinfo {author} {\bibfnamefont {K.}~\bibnamefont
  {Visscher}}, \bibinfo {author} {\bibfnamefont {M.~J.}\ \bibnamefont
  {Schnitzer}}, \ and\ \bibinfo {author} {\bibfnamefont {S.~M.}\ \bibnamefont
  {Block}},\ }\href@noop {} {\bibfield  {journal} {\bibinfo  {journal}
  {Nature}\ }\textbf {\bibinfo {volume} {400}},\ \bibinfo {pages} {184}
  (\bibinfo {year} {1999})}\BibitemShut {NoStop}%
\bibitem [{\citenamefont {Abramowitz}\ and\ \citenamefont
  {Stegun}(1964)}]{abramowitz+stegun}%
  \BibitemOpen
  \bibfield  {author} {\bibinfo {author} {\bibfnamefont {M.}~\bibnamefont
  {Abramowitz}}\ and\ \bibinfo {author} {\bibfnamefont {I.~A.}\ \bibnamefont
  {Stegun}},\ }\href@noop {} {\emph {\bibinfo {title} {Handbook of Mathematical
  Functions with Formulas, Graphs, and Mathematical Tables}}},\ \bibinfo
  {edition} {ninth dover printing, tenth gpo printing}\ ed.\ (\bibinfo
  {publisher} {Dover},\ \bibinfo {address} {New York},\ \bibinfo {year}
  {1964})\BibitemShut {NoStop}%
\bibitem [{\citenamefont {Hancock}\ and\ \citenamefont
  {Howard}(1999)}]{Hancock1999}%
  \BibitemOpen
  \bibfield  {author} {\bibinfo {author} {\bibfnamefont {W.~O.}\ \bibnamefont
  {Hancock}}\ and\ \bibinfo {author} {\bibfnamefont {J.}~\bibnamefont
  {Howard}},\ }\href@noop {} {\bibfield  {journal} {\bibinfo  {journal} {Proc.
  Natl. Acad. Sci. U.S.A.}\ }\textbf {\bibinfo {volume} {96}},\ \bibinfo
  {pages} {13147} (\bibinfo {year} {1999})}\BibitemShut {NoStop}%
\bibitem [{\citenamefont {Kolomeisky}\ and\ \citenamefont
  {Fisher}(2000)}]{Kolomeisky2000}%
  \BibitemOpen
  \bibfield  {author} {\bibinfo {author} {\bibfnamefont {A.~B.}\ \bibnamefont
  {Kolomeisky}}\ and\ \bibinfo {author} {\bibfnamefont {M.~E.}\ \bibnamefont
  {Fisher}},\ }\href@noop {} {\bibfield  {journal} {\bibinfo  {journal}
  {Physica A}\ ,\ \bibinfo {pages} {1}} (\bibinfo {year} {2000})}\BibitemShut
  {NoStop}%
\bibitem [{\citenamefont {Wang}\ \emph {et~al.}(2003)\citenamefont {Wang},
  \citenamefont {Peskin},\ and\ \citenamefont {Elston}}]{Wang2003}%
  \BibitemOpen
  \bibfield  {author} {\bibinfo {author} {\bibfnamefont {H.}~\bibnamefont
  {Wang}}, \bibinfo {author} {\bibfnamefont {C.~S.}\ \bibnamefont {Peskin}}, \
  and\ \bibinfo {author} {\bibfnamefont {T.~C.}\ \bibnamefont {Elston}},\
  }\href@noop {} {\bibfield  {journal} {\bibinfo  {journal} {J. Theor. Biol.}\
  }\textbf {\bibinfo {volume} {221}},\ \bibinfo {pages} {491} (\bibinfo {year}
  {2003})}\BibitemShut {NoStop}%
\end{thebibliography}
%\bibilographystyle{unsrt}
%\end{comment}
%%%%%%%%%%%%%%%%%%%%%%%%%%%%%%%%%%%%%%%%%%%%%%%%%%%%%%%%%%%%%%%%%%%%%%%%%%%%%%%%%%%%%%%
\clearpage
\newpage

\end{document}